\documentclass{agujournal2019}
\usepackage{xurl} 
\usepackage[inline]{trackchanges} 
\usepackage{soul}
\usepackage{tabularx}
\DeclareUnicodeCharacter{00A0}{~}

\draftfalse

\journalname{JGR: Planets}

\usepackage{amsmath}
\usepackage{booktabs}
\usepackage{longtable}
\usepackage{tabularx}
\usepackage{array}
\usepackage{caption}

\begin{document}

%
%


\title{A scaling relation for core heating by giant impacts and implications for dynamo onset}

%
%




\authors{You Zhou\affil{1,2}, Peter E. Driscoll\affil{2}, Mingming Zhang\affil{2}, Christian Reinhardt\affil{3,4},Thomas Meier\affil{3}}

\affiliation{1}{Planetary Science Research Center, College of Earth Sciences, Chengdu University of Technology, Chengdu, SiChuan, China}
\affiliation{2}{Earth and Planets Laboratory, Carnegie Institution for Science, Washington, DC, USA}
\affiliation{3}{Department of Astrophysics, University of Zurich, Winterthurerstrasse 190, CH-8057 Zurich, Switzerland}
\affiliation{4}{Physics Institute, Space Research and Planetary Sciences, University of Bern, Sidlerstrasse 5, CH-3012 Bern, Switzerland}





\correspondingauthor{You Zhou}{zhouyou06@cdut.edu.cn}
\correspondingauthor{P. Driscoll}{pdriscoll@carnegiescience.edu }



\begin{keypoints}
\item 
A giant impact can significantly heat Earth's core with an amplitude and distribution that  depends on the impact conditions

\item Giant impacts induce heterogeneous heating within the core, often leading to thermal stratification, inhibiting dynamo onset

\item Scaling relations are developed relating impact conditions with the post-collision radial temperature profile of Earth's core

\end{keypoints}

%
%

%
%


\begin{abstract}

Accretional heating of Earth's interior during formation is pivotal to its subsequent thermal and chemical evolution. In particular, impact heating of Earth's core is expected, but its amplitude and radial distribution within the core is unknown and could influence the onset of the geodynamo. The uncertainty is due, in part, to the lack of  constraints on the temperature of the interior following formation due to the difficulty of preserving a record of such a high energy environment, and the assertion that super-heating during formation would be rapidly lost through magma ocean cooling. Here we systematically investigate core heating due to giant impacts using a Smoothed Particle Hydrodynamics (SPH) code with simulations spanning a range of impact angles, velocities, and masses. From these simulations we derive a scaling relation for core heating that depends on the impact parameters and predicts the radial core temperature profile following the impact. Our findings show that a significant amount of heat is deposited into the core, with a canonical impact scenario resulting in an average core temperature increase of about 3000 K, approximately 500 K higher than that of the overlying mantle. In this case the heat distribution within the the core produces a strong thermal stratification. We use a parameterized cooling model to estimate that the core could have cooled to an adiabatic state $\sim290$ Myr after a canonical impact, which is consistent with the observed time span between the age of the Moon and evidence for an active geodynamo.
\end{abstract}

\section*{Plain Language Summary}
Understanding how the Earth cools down is a pivotal question in the study of our planet's evolutionary history. The initial thermal state of the Earth's core plays a significant role because it determines when the geomagnetic field first emerged. To study this issue, we systematically investigated core heating during giant impacts using a Smoothed Particle Hydrodynamics (SPH) code. Each simulation runs a single giant impact under specific initial conditions. These simulations covered a range of impact angles, velocities, and impactor masses. First, we found that different giant impacts cause significant differences in the Earth's core temperature. Second, the temperature distribution within the core is highly heterogeneous, with high temperatures mainly concentrated at the outermost part of the core. This leads to the natural formation of a thermally stratified structure in the core. Such a stratified structure is very stable and will not undergo convection, which may delay the initiation of the Earth's earliest magnetic field.

%
%

%


%
%
%
%

\section{Introduction}
The thermal cooling of Earth's interior from its initial state is central to better understand our planet's history. Fundamentally, Earth's earliest thermal state influences the time scales of core-mantle differentiation, magma ocean solidification, and geodynamo onset. The depth and magnitude of internal heat deposited during giant impacts is difficult to constrain from petrologic and isotopic measurements as they are mostly reset in the high energy environment of the impact.  Numerical simulation of giant impacts offers unique insight into the nature of the heating.


After the Moon-forming giant impact, the proto-Earth completed its major accretion process, with subsequent late accretion only contributing about 0.3$\%$ to 1$\%$ of Earth's mass \cite{day2016highly}. Consequently, the Moon-forming impact played a crucial role in determining Earth's initial interior thermal and compositional state. Among the various aspects of Earth's initial thermal conditions, the state of Earth's core is of particular significance, as it profoundly influences the planet's long-term thermal evolution and the generation of the geomagnetic field through dynamo action in the core. Despite its importance, knowledge of the initial state of Earth's core remains very uncertain.

Pioneering research by \citeA{nakajima2015melting} focused on the impact-induced entropy gain of the mantle, aiming to explore the melting and mixing states of Earth's mantle after the Moon-forming impact. Building on their work, \citeA{nakajima2021scaling} developed mantle melt scaling relations based on several impact parameters, providing accurate descriptions of heat distribution within the mantle and highlighting significant pressure disparities between global magma oceans and spatially confined melt pools. \citeA{carter2020energy} demonstrated that giant impacts significantly increased the internal energies of celestial bodies, influencing the degree of melting, vaporization, thermal inflation, and spin rate of the final planets. In a comprehensive study, \citeA{lock2020energy} examined the energy changes during Earth's recovery after the Moon-forming impact, emphasizing the importance of kinetic, potential, and internal energy in shaping Earth's thermal structure and evolution timescales. Recently, \citeA{marchi2023long} proposed that shock heating resulting from impact events in the  late accretion stage could heat Venus's core, subsequently affecting its long-term thermal evolution, including long-lived resurfacing of its terrain. 


In this study we develop a scaling relation for the resulting thermal profile in the core following a giant impact that depends on the impact angle, impact velocity, impactor mass and initial thermal state.  We highlight impact conditions that deposit significant heat into the core, and the depth in the core at which the heat is deposited. Finally we discuss the implications of core impact heating to the thermal and magnetic evolution of the Earth and terrestrial planets. Our results indicate that if impact angles are less than or equal to 45 degrees, giant impacts can lead to significant heating of the Earth's core, with the most extreme cases increasing the core temperature to 14,068 K. Moreover, the heterogeneous distribution of heat within the core in the radial direction will result in thermal stratification of the core, influencing the onset of the dynamo. 

A brief outline of the paper is as follows.  Section \ref{methods} describes the numerical Methods of the smooth particle hydrodynamic (SPH) code.  Section \ref{results} details the results of the simulations and the development of a scaling law that describes the radial temperature distribution in the core after the impact for a range of impact conditions.  Section \ref{discussion} discusses the implications of the simulation results in terms of the cooling time required to return the core to an adiabatic state.  Section \ref{conclusions} summarizes our main findings.  The Supplemental Information contains more details about the simulations and their interpretation.

\section{Methods}\label{methods}

\subsection{SPH model}

We utilize the Smoothed Particle Hydrodynamics (SPH) modeling approach to simulate giant impacts, which been widely used to investigate collisions during the formation of terrestrial planets, such as the Moon-forming giant impact \cite<e.g.,>[etc.] {canup2001origin,canup2012forming,cuk2012making,reufer2012hit,nakajima2014investigation,rufu2017multiple, hosono2019terrestrial, nakajima2021scaling, kegerreis2022immediate, Timpe_2023} and the Mercury-forming giant impact \cite<e.g.,>[etc.]{benz1988collisional,benz2008origin,asphaug2014mercury,chau2018forming,deng2019hypothetical,reinhardt2022forming}. Here we employ a state-of-the-art SPH code named Gasoline \cite{wadsley2004gasoline}.  This version of the code has been used in several previous studies and incorporates several improvements, including the ability to resolve sharp density contrasts at metal and silicate boundaries, and prevent significant entropy loss during the simulation. \cite{reinhardt2017numerical,chau2018forming,reinhardt2020bifurcation,reinhardt2022forming}. 
The governing equations for planetary SPH are shown in (\ref{Continuity equation}) to (\ref{mu}).

 Continuity equation
\begin{equation}
    \frac{D\rho_i}{Dt} = \sum_{j=1}^{N} m_j (v_i - v_j) \cdot \nabla_i W_{ij}
    \label{Continuity equation}
\end{equation}

 Momentum equation
\begin{equation}
    \frac{Dv_i}{Dt} = -\sum_{j=1}^{N} m_j \left(\frac{P_j}{\rho_j^2} + \frac{P_i}{\rho_i^2}\right) \cdot \nabla_i W_{ij}
    \label{Momentum equation}
\end{equation}

 Energy equation
\begin{equation}
    \frac{Du_i}{Dt} = \frac{1}{2} \sum_{j=1}^{N} m_j \left(\frac{P_j}{\rho_j^2} + \frac{P_i}{\rho_i^2}\right) v_{ij} \cdot \nabla_i W_{ij}
    \label{Energy equation}
\end{equation}

where $\rho_i$ is the density of particle $i$, $v_i$ is its velocity vector, and $e_i$ is its internal energy vector.  $W_{ij}$ is the smoothing function and $\nabla_i W_{ij}$ is the gradient of the smoothing function, $h$ is the smoothing length that defines the influence area of the smoothing function. $N$ is the number of particles in the neighborhood, and $m_j$ represents the mass of each neighboring particle to particle $i$. $P$ is the pressure and it can be computed via $P_i = \text{EoS}(\rho_i, e_i)$. To address discontinuities arising from shocks in SPH simulation, a common approach  involves incorporating an artificial viscosity term. 

 Expression for artificial viscosity (momentum part)
\begin{equation}
    \left. \frac{Dv_i}{Dt} \right|_{av} = -\sum_{j=1}^{N} m_j \Pi_{ij} \cdot \nabla_i W_{ij}
    \label{artificial viscosity (momentum part)}
\end{equation}

 Expression for artificial viscosity (energy part)
\begin{equation}
    \left. \frac{Du_i}{Dt} \right|_{av} = \frac{1}{2} \sum_{j=1}^{N} m_j \Pi_{ij} v_{ij} \cdot \nabla_i W_{ij}
    \label{artificial viscosity (energy part)}
\end{equation}

The artificial viscosity $\Pi_{ij}$ is parametrized by
\begin{equation}
\begin{cases}
    \Pi_{ij} = -\frac{\alpha c_{ij} \mu_{ij} + \beta \mu_{ij}^2}{\rho_{ij}}, & \text{for } (v_i - v_j) \cdot (r_i - r_j) > 0 \\
    0, & \text{else}
\end{cases}
\label{artificial viscosity}
\end{equation}

In the expression (\ref{artificial viscosity}), 
\begin{equation}
\mu_{ij} = -\frac{h (v_i - v_j) \cdot (r_i - r_j)}{(r_i - r_j)^2 + \eta h^2}
\label{mu}
\end{equation}

where $c_{ij}$ and $\rho_{ij}$ are the average sound speed and density, and $\alpha$ and $\beta$ define the amount of dissipation in the simulation. $r_i$ denotes the position of particle $i$. Following previous work, we set $\alpha = 1.5$ and $\beta = 3$ for the viscosity parameters \cite{canup2004simulations}. Additionally, we set $\eta = 0.01$ to prevent divergence when two particles come too close to each other.

\subsection{Initial conditions}

In our simulations both the proto-Earth and the impactor are initialized by a hydrostatic density profile. However, the pressure forces, which depend on the initial distance between particles and their density, tend to generate oscillations. To address this issue, we employ a relaxation process to ensure hydrostatic equilibrium in each SPH particle. This is achieved by applying a velocity damping method, setting the damper coefficient at 0.95 \cite<e.g.,>{zhou2021core,reinhardt2017numerical}. The resulting initial density, pressure, and temperature profiles of the proto-Earth and impactor yield low noise, with the particle's average velocity being about 1.5 m/s, satisfying the requirements of giant impact simulations (see Supplementary \textit{Figure S1}). Our simulations also utilized the latest M-ANEOS equation of state (EOS), which ensures accurate physical behavior during giant impact events under conditions of high-temperature and low-density due to the inclusion of a liquid phase \cite{stewart2020shock,reinhardt2022forming}. A Forsterite EOS table was used to model the mantle, and an pure iron EOS table for the core \cite{stewart_sarah_t_2019,sarah_t_stewart_2020}. 

Prior to the impact, both proto-Earth and the impactor are initialized with 30$\%$ core mass fraction and 70$\%$ mantle mass fraction \cite{canup2004simulations}. Both bodies are initialized with a surface temperature of 2000 K, and it extrapolates to approximately 4000 K at the proto-Earth's core center. The temperature profiles for both the core and the mantle are assumed to satisfy an adiabatic state respectively. In order to test the influence of the initial temperature we also conduct simulations with initial surface temperatures of 1000 K and 3000 K, respectively. 

The proto-Earth is represented by 500,000 particles in all cases, while the impactor is resolved by 25,000$-$150,000 particles depending on its mass. The mass of each particle is \(1.2130 \times 10^{19}\) kg. The resolution level is sufficient to accurately model the stripping of mantle material as described in previous giant impact models \cite{meier2021eos}, and it is therefore expected to adequately model energy deposition from the shock wave.
The simulation time for most collisions is about 34 hours, thought a few high-energy cases are extended to about 51 hours, consistent with previous simulations \cite<e.g.,>{zhou2021core,reinhardt2020bifurcation,reinhardt2022forming,chau2021could}. 

\subsection{Simulation parameters}
We carry out simulations with a range of impact angles, velocities, and impactor sizes to extensively analyze the heating of Earth's core due to giant impacts. Specifically, we use impact angles of $15^\circ$, $30^\circ$, $45^\circ$, $60^\circ$, and $75^\circ$ (where $0^\circ$ is head-on and $90^\circ$ is grazing); impact velocities of $1 V_{esc}$, $2V_{esc}$, and $3V_{esc}$ (where $V_{esc}$ represents the mutual escape velocity, which ranges from 9.4496 to 9.5626 km/s, attributed to the differences in their radii and masses); and impactor masses of $0.05M_{\oplus}$, $0.1M_{\oplus}$, $0.2M_{\oplus}$, and $0.3M_{\oplus}$ (where $M_{\oplus}$ is the mass of the Earth). After each impact, in order to plot the core temperature profile we recalculated the position of the center of mass of the proto-Earth's core. We removed distant particles with densities below 1$\%$ of the average density to minimize their interference with the center of mass. Following the giant impacts, mantle and core particles might overlap at the core-mantle boundary. 
In these cases, we classified the entire area as part of the core if the proportion of metal particles in this region exceeds 90$\%$.

\section{Results} \label{results}
\subsection{Core heating simulation results}

\begin{figure}
\noindent\includegraphics[width=\textwidth]{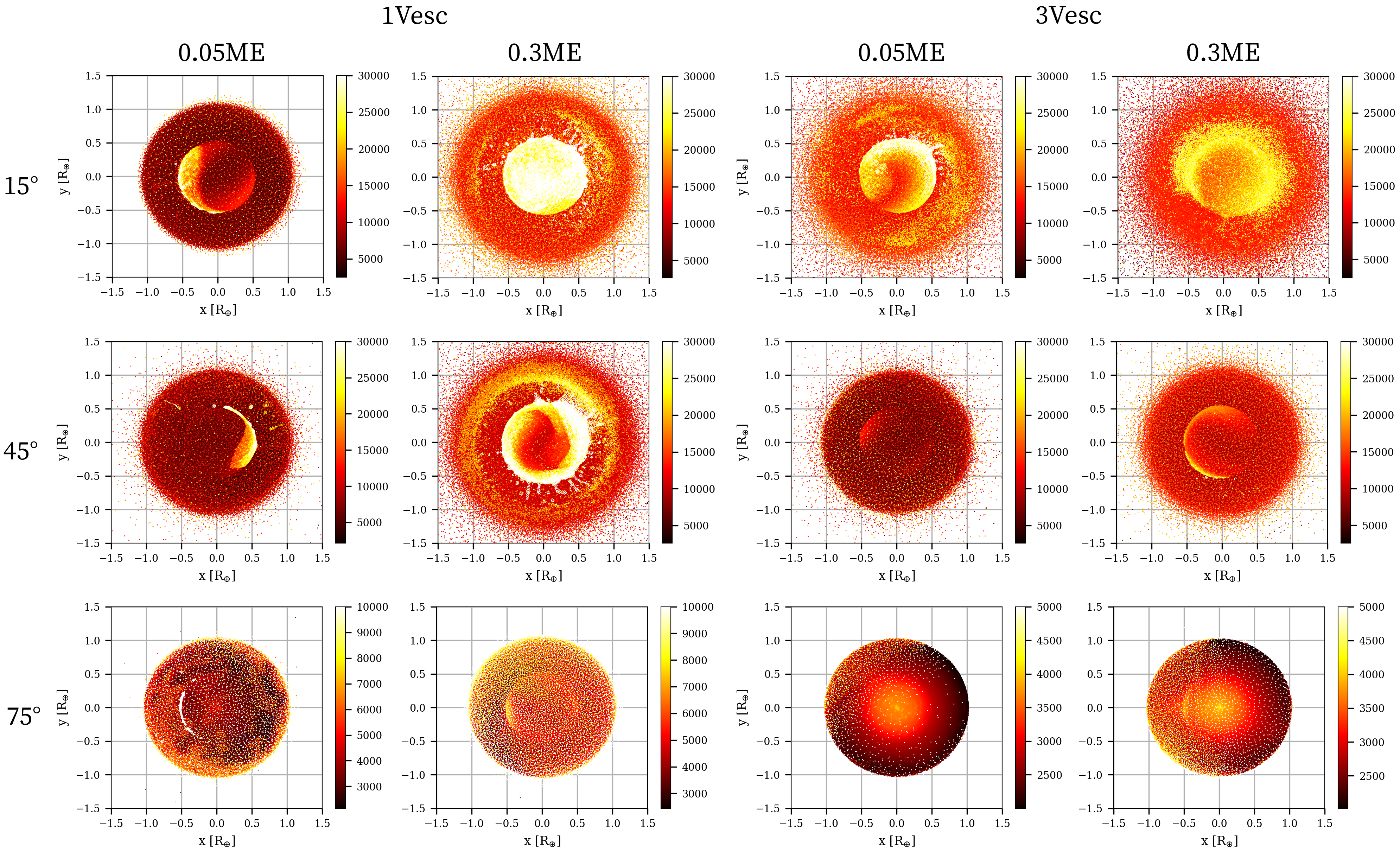}
\caption{Comparison of SPH simulations of core temperatures across diverse giant impact scenarios. Each panel displays the temperature distribution of Earth's core following the giant impacts, with units on the colorbar indicated in Kelvin (K). For the Core-merging giant impact scenarios, we've set a consistent maximum temperature threshold at 30,000K. In certain instances, it's worth noting that the temperature at the core-mantle boundary might surpass this. As for the hit-and-run scenarios, we've designated two distinct maximum temperature values: 10,000K and 5,000K. This approach is primarily designed to offer a clearer visualization of core heating results. We selected representative scenarios with the smallest impactor of $0.05M_{\oplus}$, the largest impactor of $3M_{\oplus}$, typical impact velocities at minimum ($1V_{\text{esc}}$) and maximum ($3V_{\text{esc}}$), and varying impact angles ($15^{\circ}$, $45^{\circ}$, and $75^{\circ}$). The simulation time for all instances remains consistent at 34.38 hours.}
\label{figuresample1}
\end{figure}

We conducted 60 SPH simulations that span the range of initial impact conditions described in Table \ref{table2} (see Appendix), all with an initial surface temperature of 2000 K. These simulations enabled us to calculate the average temperature rise in the core, as represented in Table \ref{table2} (see Appendix).  In the most extreme cases, the average core temperature surged by over 10,000 K, reaching a maximum of 14,068 K. In contrast, the least extreme cases resulted in a negligible increase in the average core temperature by as little as 32 K. 
 Figure \ref{figuresample1} displays the final temperature distributions in the X-Y (horizontal) plane for a subset of models, illustrating the range of final temperature distributions. From Figure \ref{figuresample1} it can be seen that core impact heating increases with low impact angle (more head-on), faster impact velocity, and larger impactor mass.

Among these impact parameters we find the impact angle is the most sensitive parameter to core heating. Collisions with impact angles less than $30^\circ$ (more head-on), typically referred to as ``core-merging giant impacts'', produce intense core heating because the impactor core experiences extreme shock heating, but is still dense enough to fall through the target mantle and merge with the target core. In contrast, collisions with impact angles exceeding $60^\circ$, typically referred to as ``hit-and-run giant impacts'', produce much less core heating (Figure \ref{figuresample5}). We find the impact velocity ranks as the second most crucial factor for core heating, with the size of the impactor as the least influential variable considered. We also conducted additional simulations at surface temperatures of 1000 K (60 cases) and 3000 K (10 cases) to explore the influence of the initial thermal state of the planet. We find that the amount of core heating is not very sensitive to initial thermal state (see below).  

To quantify the influence of each impact parameter, consider a scenario where an impactor with a mass of $0.1M_{\oplus}$ strikes at an angle of $45^{\circ}$, with a velocity of $1V_{\text{esc}}$, and a surface temperature of $2000$ K. In this case, the core's average temperature increase, $\Delta \bar{T}$, is $2498$ K. Holding all other factors constant, varying the impact angle to $15^{\circ}$ or $75^{\circ}$ results in a $\Delta \bar{T}$ of $6555$ K and $327$ K, respectively. By changing the impact speed to $2V_{\text{esc}}$ or $3V_{\text{esc}}$, the corresponding $\Delta \bar{T}$ values are $1617$ K and $2460$ K. Altering the impactor's mass to $0.05M_{\oplus}$ or $0.3M_{\oplus}$ leads to a $\Delta \bar{T}$ of $1537$ K and $6242$ K, respectively. Finally, adjusting the planetary surface temperature to $1000$ K or $3000$ K results in a $\Delta \bar{T}$ of $2966$ K and $2557$ K, respectively.

The dependence of core heating on impact parameters is similar to the findings of mantle heating in certain aspects. The simulations in \citeA{nakajima2021scaling} appear to agree that the impact angle exerts the most significant influence. Specifically, giant impacts at angles less than $30^{\circ}$  lead to a marked increase in the internal energy of the mantle and substantial magma production. 



\subsection{Scaling relations for core heating}

We develop five scaling relations for core heating that depend on the impact parameters ({Table \ref{table1}}): (1) the mean internal energy increase $\Delta \bar{E}$, (2) the mean temperature increase $\Delta \bar{T}$, (3) the CMB temperature increase $\Delta{T}_{cmb}$, (4) the mid-core temperature increase $\Delta{T}_{m}$, and (5) the central core temperature increase $\Delta{T}_c$. Among them, $ \Delta{T}_{cmb} $, $ \Delta{T}_{m} $, and $ \Delta{T}_c $ denote the temperatures at representative points in the post-impact core temperature profile. Meanwhile, $ \Delta \bar{E} $ and $ \Delta \bar{T} $ characterize the overall heating outcomes of Earth's core.


To calculate a one dimensional radial core temperature profile from all the particles in the post-impact core, we bin particles into 50 concentric spherical shells equally spaced in radius. After each giant impact simulation, we recalculate the location of the center of the core and the reposition the spherical shells. From this binned temperature profile we define the central temperature $T_c$ as the mean temperature of the inner most sphere with radius $r_i$, the midpoint temperature $T_m$ as the mean temperature of the mid-core shell within $r_{m1}$ and $r_{m2}$, and the CMB temperature $T_{cmb}$ as that in the last shell with an inner radius $r_{cmb1}$ just below the CMB and outer radius $r_{cmb}$. The change in each temperature $\Delta T$ is the mean temperature within each shell subtracted by the mean temperature of the corresponding shell from the proto-core before giant impact, as expressed by
\begin{equation}
\Delta T_c = \frac{1}{N_{c,f}} \sum_{i=0}^{N_{c,f}} T_i (r_i < r_{c}) - \frac{1}{N_{c,o}} \sum_{i=0}^{N_{c,o}} T_i (r_i < r_{c})
\label{eq:1}
\end{equation}

\begin{equation}
\Delta T_m = \frac{1}{N_{m,f}} \sum_{i=0}^{N_{m,f}} T_i (r_{m1}< r_i < r_{m2}) - \frac{1}{N_{m,o}} \sum_{i=0}^{N_{m,o}} T_i (r_{m1}< r_i < r_{m2}))
\label{eq:2}
\end{equation}

\begin{equation}
\Delta T_{cmb} = \frac{1}{N_{cmb,f}} \sum_{i=0}^{N_{cmb,f}} T_i (r_{cmb1}< r_i < r_{cmb}) - \frac{1}{N_{cmb,o}} \sum_{i=0}^{N_{cmb,o}} T_i (r_{cmb1}< r_i < r_{cmb}))
\label{eq:3}
\end{equation}
where $i$ is particle index, $N$ is number of particles, and subscripts $f$ and $o$ refer to the final and original state, respectively.
For $\Delta \bar{E}$ and $\Delta \bar{T}$, we also calculated the mean temperature and internal energy over the entire core and then subtracted the corresponding values from the proto-core before giant impact, 
\begin{equation}
\Delta \bar{E} = \frac{1}{N_{f}} \sum_{i=0}^{N_{f}} E_i (r_i < r_{cmb}) - \frac{1}{N_{o}} \sum_{i=0}^{N_{o}} E_i (r_i < r_{cmb})
\label{eq:4}
\end{equation}

\begin{equation}
\Delta \bar{T} = \frac{1}{N_{f}} \sum_{i=0}^{N_{f}} T_i (r_i < r_{cmb}) - \frac{1}{N_{o}} \sum_{i=0}^{N_{o}} T_i (r_i < r_{cmb})
\label{eq:5}
\end{equation}
where $N_o$ and $N_f$ denote the total number of SPH metal particles inside the core before and after the giant impact, and $ N_f \geq N_o $. $ T_i $ and $ E_i $ respectively represent the temperature and internal energy of each SPH metal particle within its shell.

Next we develop a new scaling relation using a non-linear least squares fitting method that links the impact parameters to the core heating associated with each giant impact. We adopt a scaling relation of the form
\begin{equation}
\Delta I = a (\cos\theta)^b \left(\frac{V_i}{V_{\text{esc}}}\right)^c \left(\frac{M_i}{M_\oplus}\right)^d
\label{eq:6}
\end{equation}


where $I$ is a quantity of interest (e.g. temperature or energy), $\theta$ represents the impact angle ($0^\circ$ is head on), $V_i$ is the impact velocity, $V_{\text{esc}}$ is the escape velocity, $M_i$ is the mass of the impactor, and $M_\oplus$ is Earth mass. Where $a$, $b$, $c$, and $d$ are the set of fit coefficients and exponents, one set for each quantity of interest. The fitting coefficients for each quantity of interest are shown in {Table \ref{table1}}.

\begin{table}
\caption{The Scaling Relation coefficients for the representative temperature values}
\label{table1}
\begin{tabularx}{\textwidth}{l *{6}{X}} 
\hline
Symbol  & $a$       & $b$     & $c$     & $d$    & R-squared & RMSE \\
\hline
$\Delta T_c$   & 8070.25 & 3.89 & 0.26  & 0.30 & 0.80 & 881.40 \\
$\Delta T_m$   & 8821.25 & 3.75 & 0.27  & 0.30 & 0.83 & 905.91 \\
$\Delta T_{\text{cmb}}$ & 22363.57 & 2.66 & -0.09 & 0.22 & 0.91 & 1450.40 \\
$\Delta \bar{E}$ & 52.41 & 3.57 & -0.30 &  0.34 & 0.89 & 2.75 \\
$\Delta \bar{T}$ & 10370.03 & 3.77 & 0.21  & 0.29 & 0.85 & 939.16 \\
\hline
\multicolumn{7}{p{\textwidth}}{$^{a}$Each R-squared and RMSE value is computed by comparing the respective fit data to the original values.} 
\end{tabularx}
\end{table}

\begin{figure}[h!]
\begin{center}
\noindent\includegraphics[width=\textwidth]{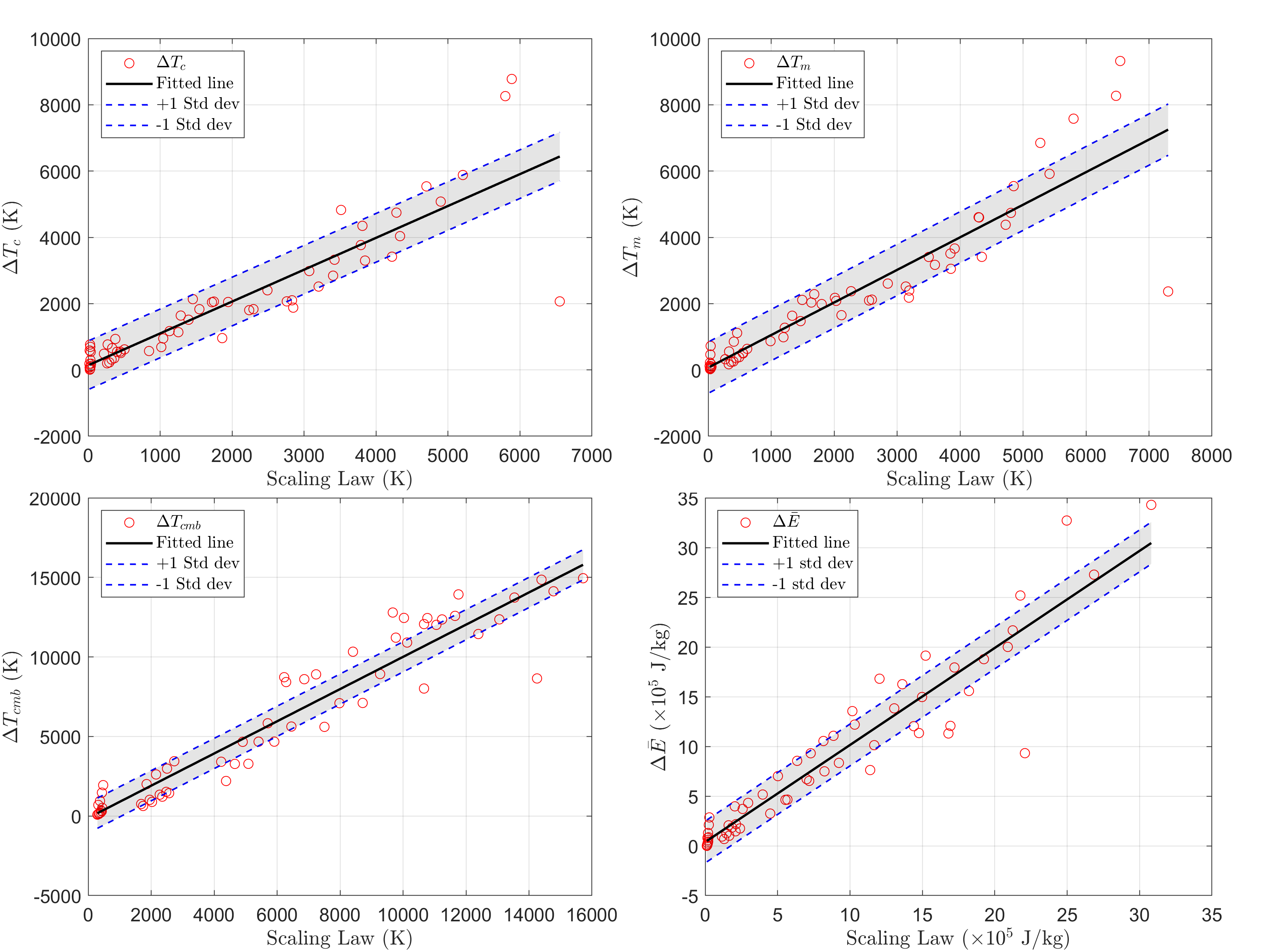}
\caption{ The relationship between the original data for quantity of interest and the values calculated using the scaling relation. The red circles denote the original data for the increase in core temperature, while the solid black line represents the fitted line. The blue dashed lines represent the range within ±1 standard deviation of the original data from the fited line. The X-axis presents the fitted values calculated using the scaling relation as described by \textit{Equation \ref{eq:6}}. The fitted parameters used for the four panels are detailed in \textit{Table \ref{table1}}. The Y-axis displays the original temperature increase at representative positions within Earth's core. }
\label{figuresample2}
\end{center}
\end{figure}

{Figure \ref{figuresample2}} compares the scaling relation in (\ref{eq:6}) to the simulation results for four of the quantities of interest, which includes all impact scenarios with an initial surface temperature set to 2000 K. Overall, the quality of the scaling relation is satisfactory, given that all the R-squared values for fitting equations in {Table \ref{table1}} exceed 0.8. The R-squared above 0.7 is typically considered an acceptable fitting outcome. Within the range where the distance between the fit line and the original data is less than one standard deviation, our scaling relation can fit the majority of cases well. {Figure \ref{figuresample2}} demonstrates a superior fit for the outer layer of the Earth’s core compared to the inner layer. Among the three representative locations on the Earth’s core temperature profile, the fit is best for $ \Delta T_{\text{cmb}} $, followed by $ \Delta T_m $, and lastly for  $ \Delta T_c $. The deviation in these fits is especially notable in scenarios with significant temperature spikes, as evident from the fit results of  $ \Delta T_c $ and $ \Delta T_m $ in {Figure \ref{figuresample2}}, which are particularly violent giant impacts. These intense collisions have a more pronounced relative influence on the inner-most core compared to the outer-most core, thereby reducing the overall goodness of fit for $ \Delta T_m $ and $ \Delta T_c $. Figures S7-S10 in the Supplementary also show that cases with poorer fits are high-energy collisions with smaller impact angles, which have a significant impact on \(T_m\) and \(T_c\).

We examine the worst fitting instance as a representative case. In this scenario, the simulated increase in the core center temperature is $2067\, \text{K}$, while the fitted model indicates $6305\, \text{K}$. Similarly, the simulated rise in mid-core temperature is $2367\, \text{K}$, but the fitted model suggests a value of $6918\, \text{K}$. The initial conditions for this instance signify a violent collision with an impact angle of $15^\circ$, a velocity of $3V_{\text{esc}}$, and an impactor's mass of $0.3 M_{\oplus}$. This violent giant impact significantly influences the inner-most core temperature, which contrasts with the effects of other cases. The fit quality for the average internal energy $ \Delta \bar{E} $ is more accurate than that of the average temperature $ \Delta \bar{T} $. Overall, our fitting for outcomes of extremely violent collisions is poor, but such collisions are also very rare. By excluding these cases, the effectiveness of our fitting will further improve.



As a cross-check, we refered to a situation described by \cite{marchi2023long}, where the impact angle is $45^\circ$, the impact velocity is 25 km/s, and the impactor mass is $0.01 M_{\oplus}$. From this data, we derive its $\Delta T_{\text{cmb}}$ to be 2974 K. Impressively, this result aligns closely with those of \citeA{marchi2023long}. This suggests that our scaling relation can accurately predict the core heating result of giant impacts during the late stage of terrestrial planet formation.



\subsection{Core thermal stratification}

\begin{figure}\begin{center}
\noindent\includegraphics[width=0.8\textwidth]{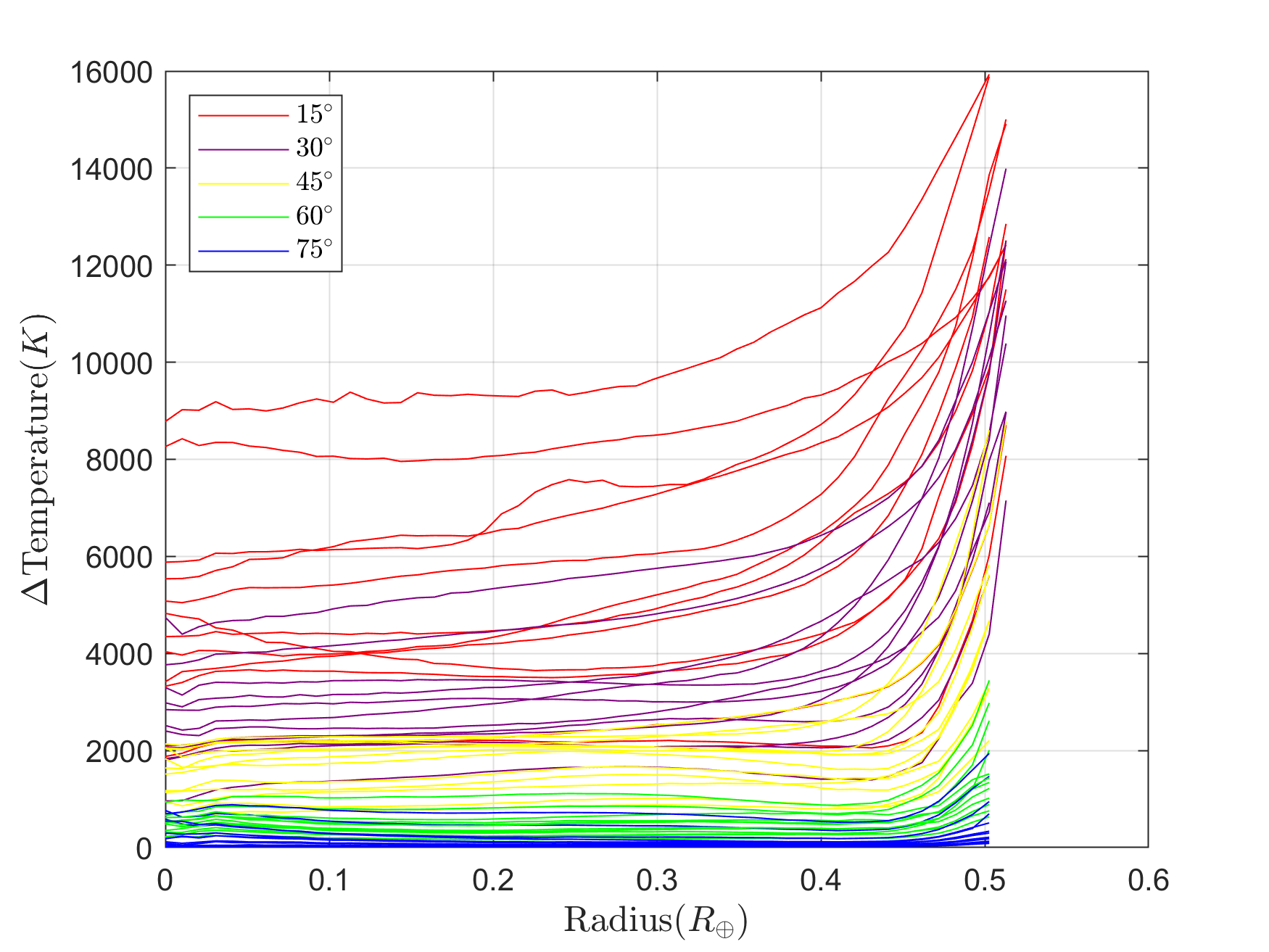}
\caption{ The change in core temperature obtained from simulations against radius. The initial surface temperature is 2000K and different colors represent the impact angle. Each distinct line represents a different simulation, totalling to 60 simulations. The region from the center of the core to the core-mantle boundary is segmented into 50 shells. Detailed criteria for the determination of the core-mantle boundary can be found in the Methods section (section 2). 
}
\label{figuresample5}
\end{center}
\end{figure}

\begin{figure}
\noindent\includegraphics[width=\textwidth]{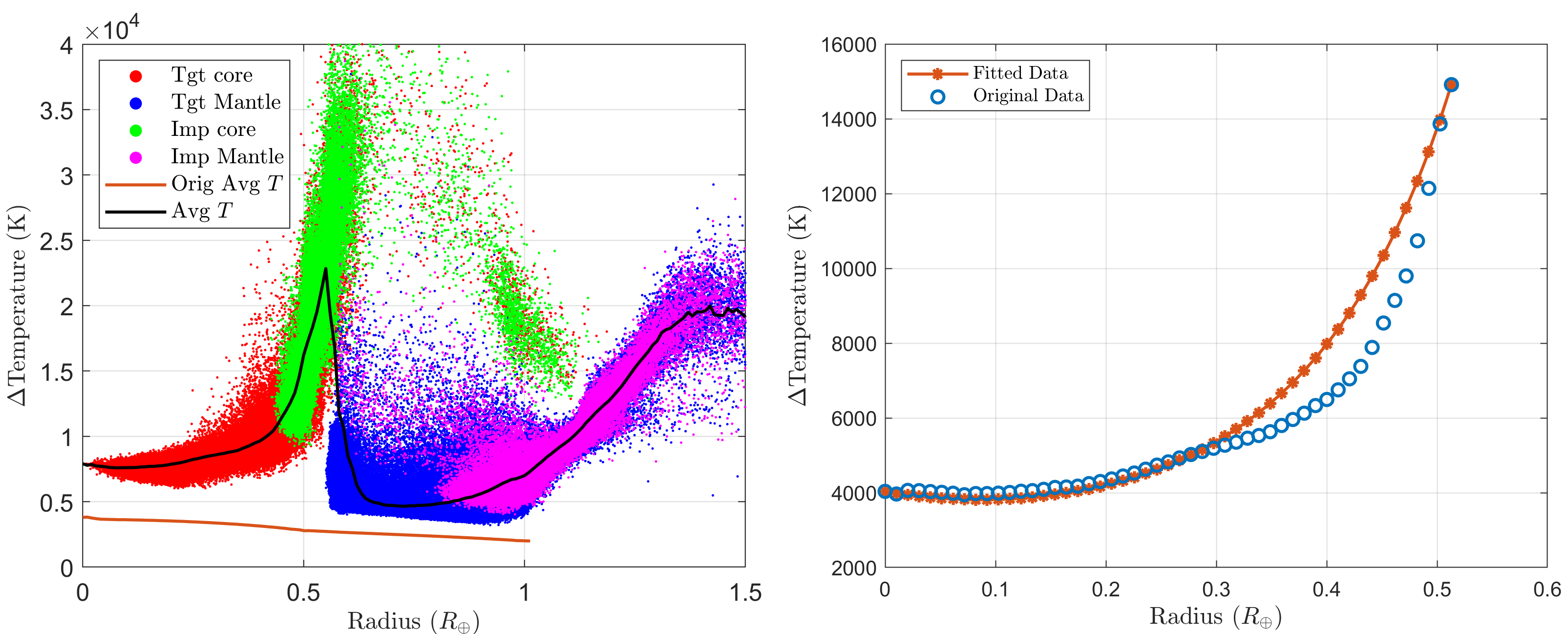}
\caption{The temperature increase of all particles against radius after the collision, and the contrast between the original data and the fitted data. This particular case was initialized with an impact angle of $15^\circ$, an impact velocity of $1V_{\text{esc}}$, and an impactor mass of $0.2M_{\oplus}$, resulting in a $\chi^2$ value of 0.46. The left panel displays the post-impact temperature distribution across all particles against the radius, and also provides the average temperature profiles before and after the collision. The right panel, on the other hand, contrasts the original data (showing the increase in core temperature from this giant impact) with the fitted data derived using Equation \ref{eq:7}. The procedure for identifying the core-mantle boundary is detailed in the Methods section (section 2).}
\label{figuresample3}
\end{figure}

\begin{figure}\centering
\noindent\includegraphics[width=0.7\textwidth]{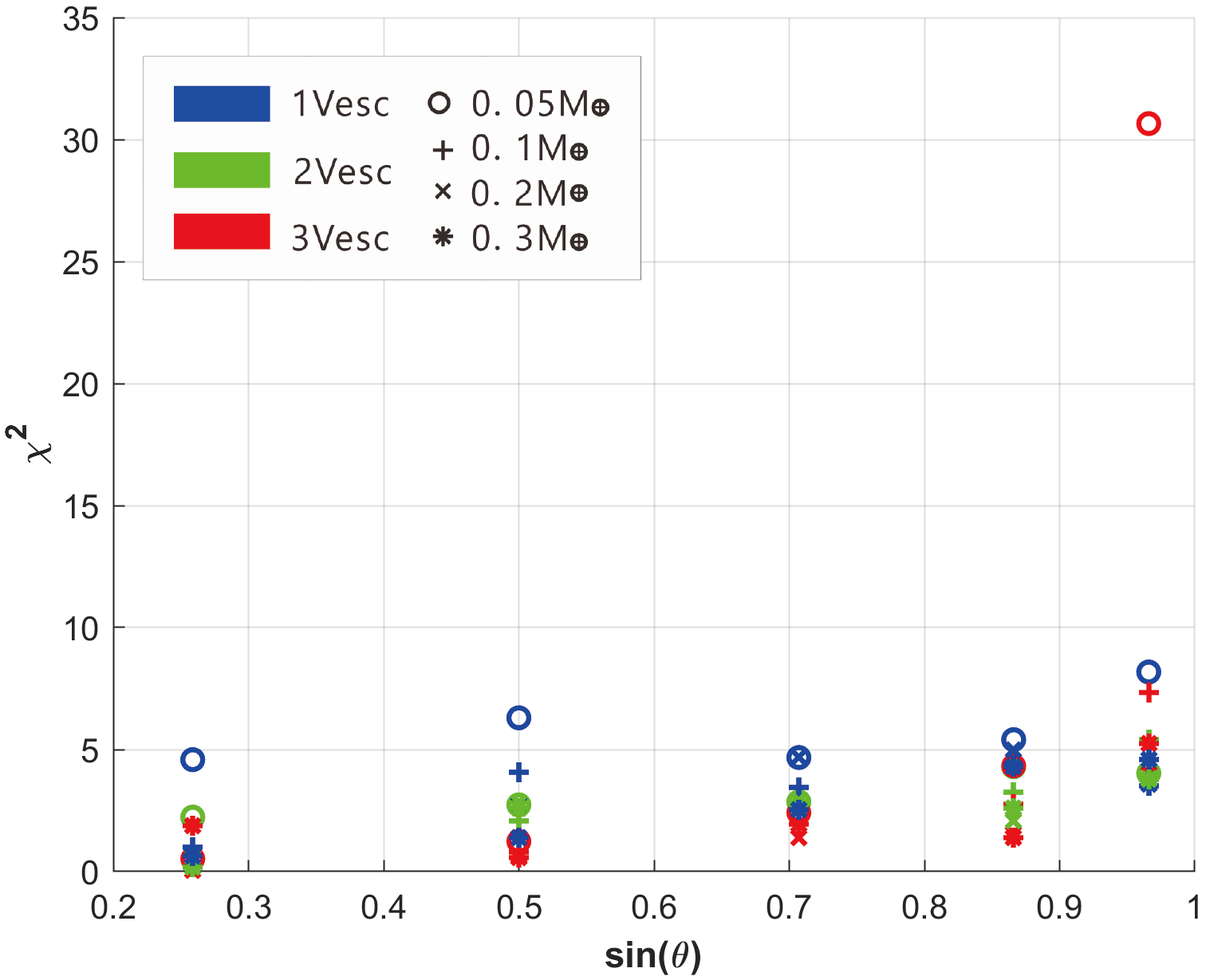}
\caption{The $\chi^2$ values of all instance run results. The expression for the error is given by Equation \ref{eq:12}. All $\chi^2$ values have been normalized by the initial temperature of the center of mass of the core. A smaller $ \chi^2 $ value represents a better goodness of fit. The x-axis represents the sine values of the impact angles, which are $0.2588$, $0.5000$, $0.7071$, $0.8660$, and $0.9659$, respectively. These correspond to angles of $15^\circ$, $30^\circ$, $45^\circ$, $60^\circ$, and $75^\circ$. }
\label{figuresample4}
\end{figure}

The distribution of impact heating in the core is crucial because it  results in a thermal stratification that will inhibit core convection and dynamo action.  
In nearly all of our simulations impact heating tends to concentrate near the CMB, as illustrated in {Figure \ref{figuresample5}} and {Figure \ref{figuresample3}}. Consequently, after a giant impact the temperature below the CMB can be several times hotter than at the center of the core as depicted in  {Figure \ref{figuresample5}} and  {Figure \ref{figuresample4}}.

This stratification arises because there is a relatively minor temperature increase in the center of the core, predominantly driven by compressional and shock heating during the impact process. In contrast, the temperature rise near the CMB is predominantly influenced by core particles from the impactor (See the left panel of {Figure \ref{figuresample3}}). We find that the particles in the impactor's core experience extreme shock heating, heating up several times more than the target core. But these impactor core particles are still denser than the mantle material, even at such extreme temperatures due to their relatively low thermal expansivity, so the impactor core sinks to the target core.  The impactor core particles then pond at the top of the target core because they are significantly hotter and less dense than the target core, similar to the findings of \citeA{landeau2016core}.

Additionally, metal particles tend to heat up more than the silicate because the specific heat capacity of metal is about $\sim 60\%$ that of silicate, meaning that uniform heating would increase the metal temperature more than the silicate. This difference contributes to the greater temperature increase in the outermost core compared to the lowermost mantle. We discuss the implications of this stratification for the geodynamo below.



\subsection{Fitting of the post-impact core temperature profile}

In addition to the total core heat gain, we can construct a radial temperature profile $T(r)$ from the scaling quantities. We find the following radial function to adequately describe the change in core temperature profiles following a giant impact,
\begin{equation}
\Delta T(r) = \Delta T_{cmb} \exp\left(\frac{R_{cmb}-r}{H_1} + \left(\frac{R_{cmb}-r}{H_2}\right)^2\right)
\label{eq:7}
\end{equation}
where $H_1$ and $H_2$ are length scales associated with the depth of impact heating. $H_1$ and $H_2$ can be written in terms of the quantities of interest by evaluating (\ref{eq:7}) at the core-mantle boundary $R_{cmb}$, mid-core radius $R_m$, and center of the core $R_c$, giving
\begin{equation}
\ln\left(\frac{\Delta T_c}{\Delta T_{\text{cmb}}}\right) = \frac{R_{cmb}}{H_1} + \left(\frac{R_{cmb}}{H_2}\right)^2
\label{eq:8}
\end{equation}
\begin{equation}
\ln\left(\frac{\Delta T_m}{\Delta T_{\text{cmb}}}\right) = \frac{R_{cmb}}{2H_1} + \left(\frac{R_{cmb}}{2H_2}\right)^2
\label{eq:9}
\end{equation}
where $R_m=R_{cmb}/2$ is used.
Then combining (\ref{eq:8},\ref{eq:9}) we can solve for $H_1$ and $H_2$ as 
\begin{equation}
H_1 = \frac{R_{cmb}}{4\ln\left(\frac{\Delta T_m}{\Delta T_{\text{cmb}}}\right) - \ln\left(\frac{\Delta T_c}{\Delta T_{\text{cmb}}}\right)}
\label{eq:10}
\end{equation}

\begin{equation}
H_2 = \frac{R_{cmb}}{\sqrt{2\ln\left(\frac{\Delta T_c}{\Delta T_{\text{cmb}}}\right) - 4\ln\left(\frac{\Delta T_m}{\Delta T_{\text{cmb}}}\right)}}
\label{eq:11}
\end{equation}

Finally, using the heating scaling relation in (\ref{eq:6}) and radial function defined in (\ref{eq:7}-\ref{eq:11}) we can calculate the core heating profile for arbitrary impact parameters. An example of such a temperature profile is shown in {Figure \ref{figuresample3}}.

To quantify the goodness of fit of our proposed exponential fitting function given by (\ref{eq:7}), we calculate the difference between the actual data and the fitted values using a $\chi^2$ formulation,
\begin{equation}
\chi^{2} =\sum_{j=0}^{N_j}\left(\frac{(\Delta T(r_j)-\Delta T_{fit}(r_j))^2}{\Delta T_{fit}(r_j)}\right)
\label{eq:12}
\end{equation}
where $ \Delta T(r_j) $ represents the temperature increase in the $ j^{\text{th}} $ radial shell, $ \Delta T_{\text{fit}}(r_j) $ denotes the predicted value from the scaling relation in (\ref{eq:6}), and $ N_j $ is the number of radial shells in the core. We normalize the $ \chi^2 $ value by the predicted value $ \Delta T_{\text{fit}}(r_j) $. Thus, a smaller \( \chi^2 \) value indicates a better goodness of fit. When \( \chi^2 = 0 \), it represents a perfect fit.

The thermal profile for each model is calculated using \eqref{eq:6} through \eqref{eq:11} and the subsequent $ \chi^2 $ value is computed  to assess the goodness of fit of the scaling relation. Our model temperature profile in (\ref{eq:7}) is designed to intersect the simulation data at three radii, which are the center $ T_c $, the midpoint $ T_m $, and the CMB $ T_{\text{cmb}} $, as demonstrated in the right panel of {Figure \ref{figuresample3}}. {Figure~\ref{figuresample4}} illustrates the $ \chi^2 $ values for each instance. Taking the simulation in {Figure \ref{figuresample3}} as an example, where the initial conditions are an impact angle of $15^\circ$, an impact velocity of $1V_{\text{esc}}$, and an impactor size of $0.2M_{\oplus}$. The $\chi^2$ value for this simulation stands at 0.46, with the maximum deviation between the original and fitted data recorded at 1908.53 K. 
Across all datasets, the average $\chi^2$ value is 3.3598, with an average maximum deviation of 1453 K. Both {Figure \ref{figuresample3}} and {Figure \ref{figuresample4}} indicate that our exponential fitting formula effectively captures the radial thermal distribution of the core across a broad range of impact parameters.

\section{Discussion}\label{discussion}

Earth's post-giant impact thermal state can be considered the starting point of its secular thermal and chemical evolution. This study demonstrates that under a wide range of impact conditions, Earth's core experiences significant heating, both from the impact itself and from the merging of the cores of the impactor and the target. Based on our simulations, all post-impact core temperature profiles increase with radius and exhibit thermal stratification. These scenarios imply that post-impact the core will be purely conductive and not convecting, inhibiting dynamo action for some time after the impact. Lateral temperature variations are produced in the  core as well ({Figure \ref{figuresample1}}), but their effect on core convection and the dynamo is expected to be small. Here we explore the implications of these scenarios in more detail and discuss some of the numerical challenges of the SPH model.


\subsection{Core Cooling Time}

\begin{figure}\centering
\noindent\includegraphics[width=0.8\textwidth]{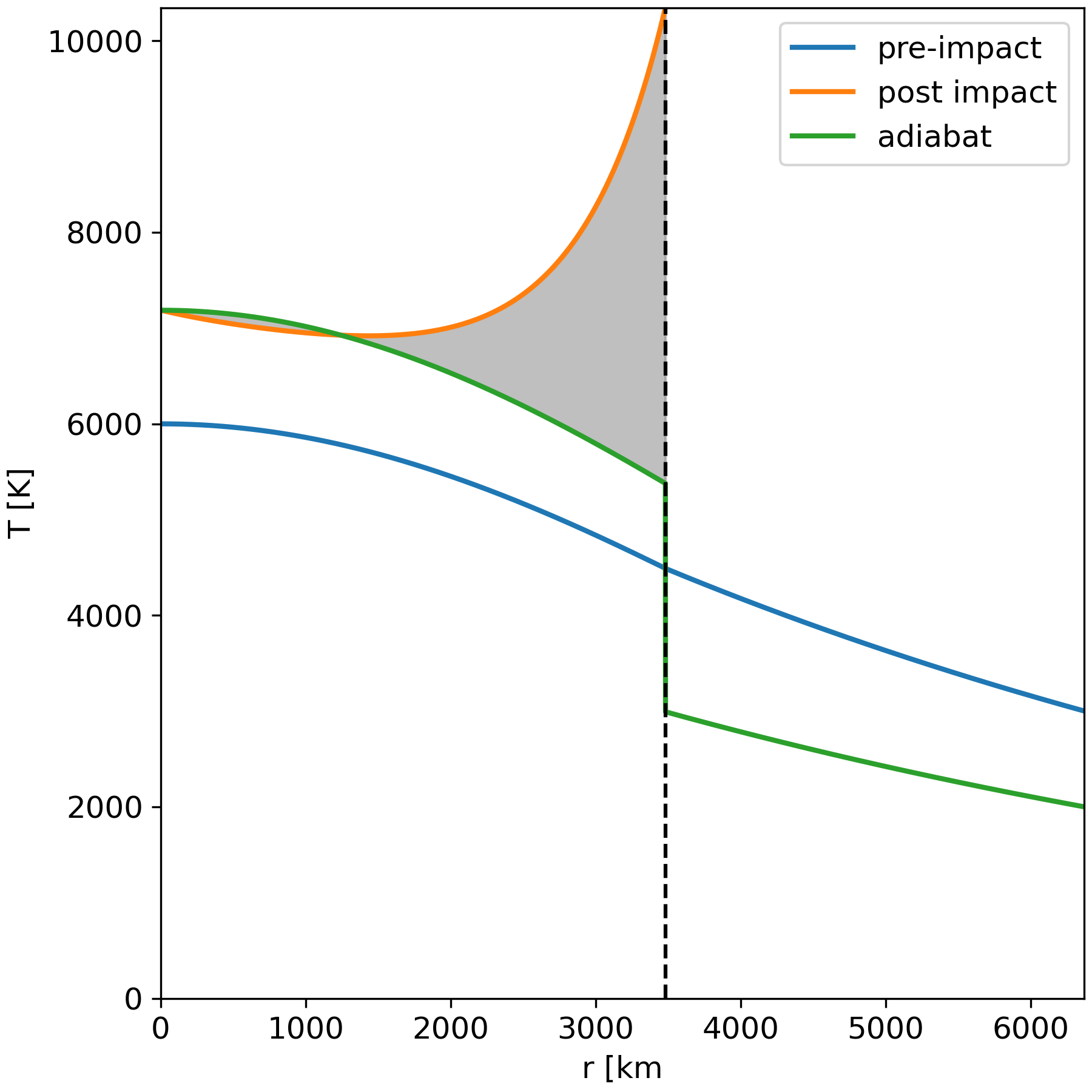}
\caption{Illustration of core super-heating during a canonical impact scenario \protect\cite{canup2004simulations}.  Comparison of the initial temperature profile in the mantle and core prior to the impact, the core temperature profile post-impact (impact parameters $\theta=46.9^\circ$, $V_i=V_{esc}$, and $M_i=0.15 M_\oplus$), and adiabatic core and mantle temperature profiles anchored to the central core temperature following the impact (see legend).  The super-heat is shaded in gray.  The CMB is denoted by a vertical dashed line.
}
\label{superheat}
\end{figure}

To initiate core convection and dynamo action the excess ``super-heat'' in the outer-most core must be transported to the mantle until an adiabatic temperature profile is reached. Figure \ref{superheat} shows the three relevant temperature profiles: the initial temperature profile in the mantle and core prior to the impact, the temperature profile post-impact, and  finally adiabatic core and mantle temperature profiles. These profiles describe the cooling process of a canonical scenario, characterized by impact parameters $\theta=46.9^\circ$, $V_i=V_{\text{esc}}$, and $M_i=0.15 M_\oplus$\cite{canup2004simulations}. The proto-Earth's surface temperature is set at 3000K. 

Given the magnitude of the giant impact heating we make the simplifying assumption that the ``super-heat'' in the core is much larger than other sources of heat (radiogenic and latent heat) so that the heat coming out of the core $Q_{CMB}$ is balanced by the heat loss rate $dE_c/dt$,
\begin{equation}
    \frac{d E_c}{d t} \approx \frac{\Delta E_c}{\Delta t} = -Q_{CMB} 
    \label{DeltaE_c}
\end{equation}
where the heat loss rate has been approximated by the amount of super-heat $\Delta E_c$ (in units of J) released over time $\Delta t$.
The time needed to cool the core to an adiabatic state is $\Delta t=t_{ad}-t_0$, where $t_{ad}$ is the time the adiabatic state is reached and $t_0$ is the time after the impact when the core and mantle have separated and the planet begins to cool.  

The amount of super-heat that needs to be removed is 
\begin{equation}
    \Delta E_c = \rho_c c_{p,c} \int_0^{R_c} (T_{imp}(r)-T_{ad}(r)) dV
\label{E_c}\end{equation}
where $\rho_c$ is core density, $c_{p,c}$ is core specific heat, $T_{imp}(r)=T_{init}(r)+\Delta T (r)$ is the core temperature profile after the impact, $T_{init}(r)$ is the initial core temperature before the impact, $\Delta T(r)$ is the heat added during the impact from (\ref{eq:7}), and $T_{ad}(r)$ is the final adiabatic temperature profile of the convecting core,
\begin{equation}
    T_{ad}(r) = T_{c}\exp\left[- \frac{r^2}{H_{c}^2}\right]
\end{equation}
where $H_{c}=\sqrt{2c_{p,c} R_c/(\alpha_c g_c)}$ is the core adiabatic scale height, $\alpha_c$ is core thermal expansivity, gravity is linear with radius $g(r)=g_cr/R_c$, and $g_c$ is gravity at the CMB (Table \ref{cooling_table}).  
Integration of (\ref{E_c}) gives
\begin{equation}
    \Delta E_c = 4 \pi \rho_c c_{p,c}  \left( \Delta T_{cmb} G_1 + T_{cmb,init} G_2 - T_{c,ad} G_3 \right) 
\label{DeltaE_c}\end{equation}
where the geometric integrals 
\begin{eqnarray}
    G_1 &=& \int_0^{R_c} r^2 \exp \left[ \frac{R_c-r}{H_1} + \left( \frac{R_c-r}{H_2}\right)^2 \right] dr \\
    G_2 &=& \int_0^{R_c} r^2 \exp \left[ - \left( \frac{r^2-R_c^2}{H_c^2}\right) \right] dr \\
    G_3 &=& \int_0^{R_c} r^2 \exp \left[ - \frac{r^2}{H_c^2}\right] dr
\end{eqnarray}
are integrated numerically using the \textit{quad} function in the \textit{python} module \textit{scipy}.

The assumption that the super-heat exceeds other sources of heat allows us to balance the CMB heat flow with the mantle surface heat flow $Q_{surf}$.  
The cooling time $\Delta t$ from (\ref{DeltaE_c}) then becomes
\begin{equation}
    \Delta t = -\frac{\Delta E_c}{Q_{surf}}~
\label{Deltat}\end{equation}

To estimate the heat flow at the surface of the mantle we use equation (8) from \cite{driscoll2014thermal} which assumes the thermal boundary layer at the top of the mantle is critical for convection,
\begin{equation}
    Q_{surf} = A_{surf} k_m \left(\frac{\alpha_m g_m}{Ra_c \kappa_m}\right)^\beta \Delta T_m^{\beta+1} \nu_m^{-\beta}
    \label{Q_surf}
\end{equation}
where $A_{surf}$ is surface area, $k_m$ is thermal conductivity, $\alpha_m$ is thermal expansivity, $g_m$ is gravity, $\kappa_m$ is thermal diffusivity, $Ra_c$ is the critical Rayleigh number for convection, $\nu$ is viscosity, $\beta$ is a cooling exponent, and $\Delta T_m=T_{UM}-T_{surf}$ is the temperature across the thermal boundary layer (Table \ref{cooling_table}). Temperature-dependent viscosity is described by an Arrhenius law
\begin{equation}
    \nu=\nu_0 \exp{\left[\frac{A_\nu}{R_g T_{UM}}\right]}
\end{equation}
where $\nu_0$ is a reference viscosity, $A_\nu$ is activation energy, and $R_g$ is the gas constant (Table \ref{cooling_table}).

Rather than compute a full thermal evolution of the mantle and core, we make the simplifying assumption that the upper mantle magma ocean will quickly cool down to a partially solid state in $\sim 10$ Myr \cite{solomatov2015,monteux2016} and that the core super-heat will be lost after this initial phase.  During this initial solidification most of the secular cooling will occur in the mantle, and little of the core heat will be lost.  
The temperature at the end of this initial cooling is assumed to be equal to an approximate upper mantle peridotite liquidus temperature of $T_{UM}\simeq 2000$ K \cite{monteux2016,herzberg1996}.   
The adiabatic mantle profile plotted in Figure \ref{figuresample6} is \begin{equation}
    T_{ad,m}=T_{UM} \exp \left[ - \frac{r-R_s}{H_m}\right]
\end{equation}
where the mantle adiabatic scale height is $H_m=c_{p,m}/(\alpha_m g_m)$, $c_{p,m}$ is mantle specific heat, and gravity in the mantle $g_m$ is assumed to be uniform (Table \ref{cooling_table}).



\begin{figure}[h!]\centering
\noindent\includegraphics[width=0.6\textwidth]{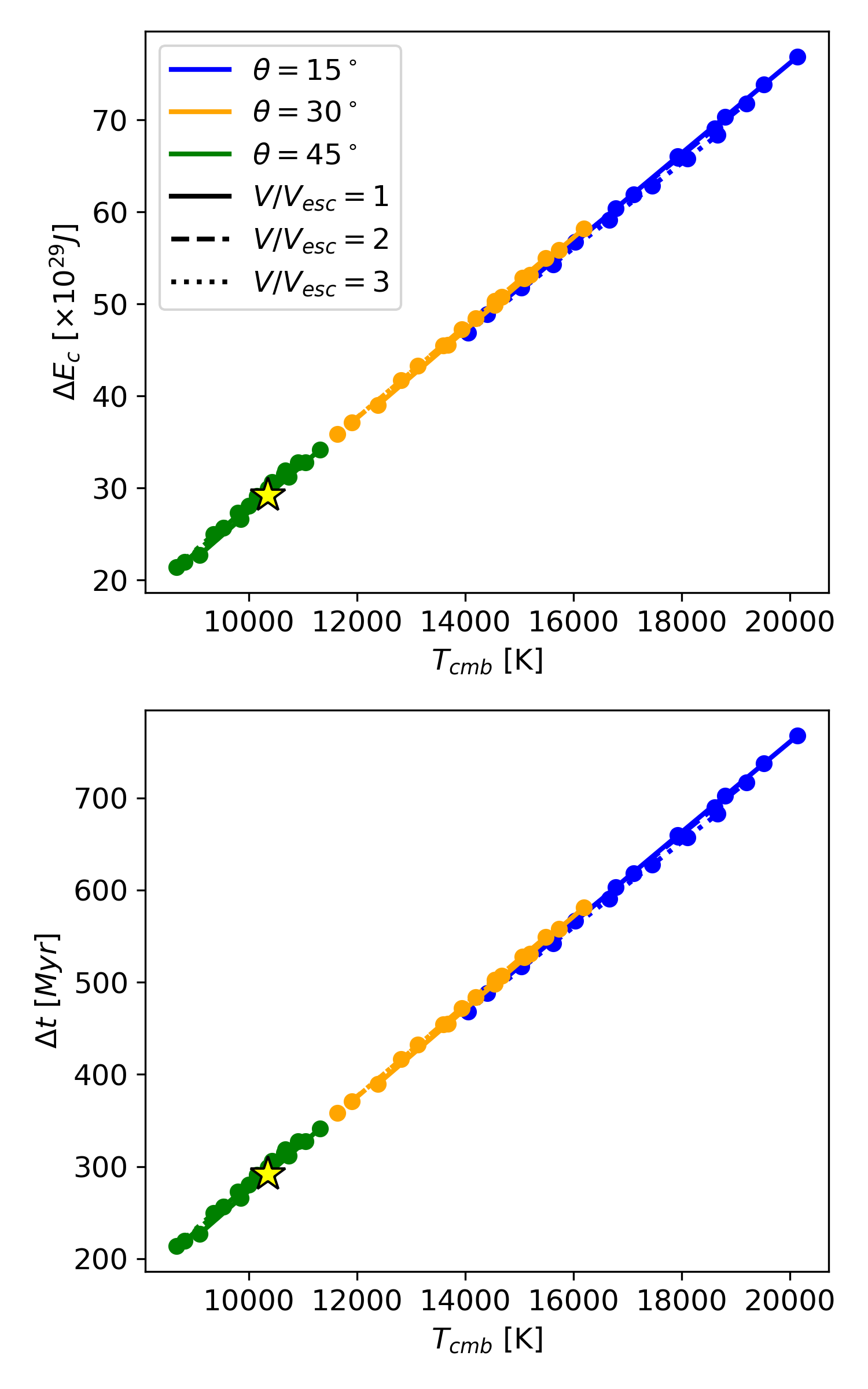}
\caption{(a) Post-giant impact core super-heat $\Delta E_c$ calculated from (\ref{DeltaE_c}) versus post-impact $T_{cmb}$ (CMB temperature at time $t_0$) for a range of impact parameters (see caption).  For each set of impact angle $\theta$ (color) and impactor velocity $V_i/V_{esc}$ (linestyle), impactor masses $M_i/M_{\oplus}$ are shown from $0.05$ to $0.3$ in increments of $0.05$ (circles).  The canonical impact ($\theta=45^\circ$, $V_i=V_{esc}$, and $M_i=0.15 M_\oplus$) is shown as a star.
(b) Time to remove core super-heat $\Delta t$ from (\ref{Deltat}).  Symbols are the same as in (a).}
\label{cooling}
\end{figure}

The super-heat $\Delta E_c$ calculated from (\ref{DeltaE_c}) is shown in Figure \ref{cooling}a versus $T_{cmb}$ (CMB temperature at time $t_0$) for a range of impact parameters (see caption).  The amount of core super-heat depends mainly on impact angle, with the nearly head-on cases $\theta=15^\circ$ producing the largest stratification.
Figure \ref{cooling}b shows the core super-heat cooling time $\Delta t$ for the same cases shown in Figure \ref{cooling}a.  The cooling times range from $200-700$ Myr over the impact scenarios considered.
Is this consistent with Earth's core cooling the core down to an adiabatic state by the time the geodynamo is known to have been active?  Given the recent revisions to the age of the Moon of $4460~\pm31$ Ma \cite{greer2023} and the earliest claims of paleomagnetic evidence supporting the existence of the geodynamo around $\sim4200$ Ma \cite{tarduno2020}, core super-heat should be removed over $\sim 260$ Myr.  For a canonical impact we find a cooling time of $290$ Myr, which is within the uncertainty of the difference in ages listed above. 
These results also imply that the onset of the dynamo could have been delayed by an additional $\sim500$ Myr with a more direct or more energetic impact.

\begin{table}
\caption{Core cooling model parameters.  All values from \protect\citeA{driscoll2014thermal} unless otherwise noted.}
\begin{tabularx}{\textwidth}{l| l | l | l}
Parameter  & Value       & Units & Description \\
\hline
$A_\nu$ & $3\times10^5$ & [J mol$^{-1}$] & viscosity activation energy \\
$\alpha_c$ & $1.4\times 10^{-5}$ & [K$^{-1}]$ & core thermal expansivity \\
$\alpha_m$ & $1.8\times 10^{-5}$ & [K$^{-1}]$ & upper mantle thermal expansivity \\
$\beta$ & $1/3$ & - & mantle cooling exponent \\
$c_{p,c}$ & $840$ & [J kg$^{-1}$K$^{-1}$] & core heat capacity \\
$c_{p,m}$ & $1265$ & [J kg$^{-1}$K$^{-1}$] & mantle heat capacity \\
$g_c$ & $10$ & [m s$^{-2}$] & CMB gravity \\
$g_m$ & $9.8$ & [m s$^{-2}$] & mantle gravity \\
$H_c$ & $6463$ & [km] & core adiabatic scale height \\
$H_m$ & $7171$ & [km] & mantle adiabatic scale height \\
$k_m$ & $4.2$ & [W/m/K$]$ & mantle thermal conductivity \\
$\kappa_m$ & $1\times10^{-6}$ & [m$^2$s$^{-1}$] & upper mantle thermal diffusivity \\
$\nu_0$ & $1\times 10^7$ & [m$^2$s$^{-1}$ & upper mantle reference viscosity \\
$\rho_c$ & $1\times 10^{5}$ & [kg m$^{-3}$] & core density \\
$Ra_{crit}$ & $660$ & - & critical Rayleigh number \\
$R_g$ & $8.314$ & [J mol$^{-1}$K$^{-1}$] & gas constant \\
$T_{surf}$ & $300$ & [K] & surface temperature \\
\end{tabularx}
\label{cooling_table}
\end{table}

\subsection{Lateral thermal heterogeneity}
Our giant impact simulations suggest that the heating distribution within the core exhibits pronounced lateral heterogeneity (as shown in {Figure \ref{figuresample1}}).
This is similar to previous studies that find large impacts during the late accretion stage could lead to thermal asymmetry in the core and mantle \cite{watters2009thermal,nakajima2021scaling}.  As discussed above, radial convection will be inhibited in the core until the super-heat is removed, which can take several 100 Myr. 
However, horizontal temperature perturbations are expected to rapidly homogenize because the mantle and core are well above their liquidus temperatures, implying low viscosities and rapid mixing rates.  

\begin{figure}[h!]\centering
\noindent\includegraphics[width=0.8\textwidth]{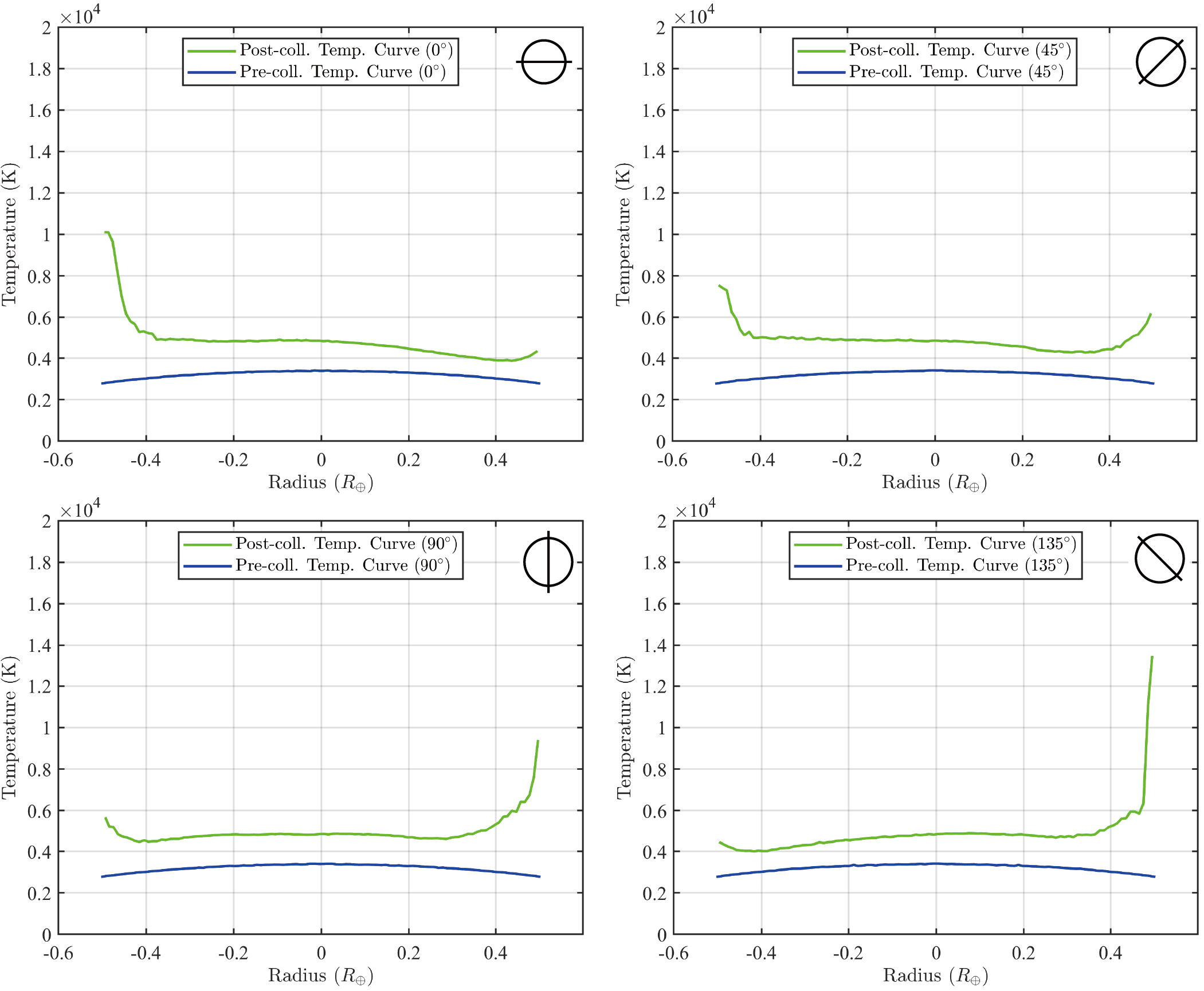}
\caption{The post-impact temperature profiles of the core in four directions are shown, corresponding to angles of 0\textdegree, 45\textdegree, 90\textdegree, and 135\textdegree\ relative to the X-axis. Blue lines indicate the temperature profiles along these directions before the impact, while green lines represent the profiles after the impact. The values of the temperature profiles are the average temperatures of particles within a 0.1R$_{\oplus}$ radius range around each profile line. The legends in the top right corners of each panel denote the direction of each temperature profile. The procedure for identifying the core-mantle boundary is detailed in the Methods section (section 2).}
\label{thermal asymmetry}
\end{figure}

To illustrate the amplitude of lateral heating following a giant impact, we created four cross-sectional lines through the post-impact core at 0\textdegree, 45\textdegree, 90\textdegree, and 135\textdegree. Particles within a 0.1R$_{\oplus}$ radius of each section line were used to represent the temperature values for that line. {Figure \ref{thermal asymmetry}} displays a typical example, showing the temperatures along these four sections post-impact. In this instance, the impactor's mass is 0.1M$_{\oplus}$, the impact velocity is 1V$_{\text{esc}}$, and the impact angle is 45\textdegree. In this case the post-impact core temperature varies laterally from a high of 13{,}472 K to a low of 4{,}347 K. The Supplemental {Figures S3-S6} further demonstrate this phenomenon with impact angles of 15\textdegree, 30\textdegree, 60\textdegree, and 75\textdegree, all of which display a similar heterogeneous heat distribution in the lateral direction, consistent with the 45\textdegree  results. 

Therefore, the lateral temperature variations did indeed occur, but they would be short-lived. \citeA{arkanihamed2010} has demonstrated that any pre-existing core temperature variations would be quickly homogenized following a giant impact due to the low viscosity of the liquid core. The lateral equilibration time of an inviscid fluid parcel of width $D$ and lateral temperature difference $\Delta T$ can be estimated by the Brunt-Vaisala frequency,
\begin{equation}
    N^2=-\frac{g}{\rho_0} \frac{d\rho}{dy} = \frac{\alpha_c g_c \Delta T}{D}
    \label{N2}
\end{equation}
where $d\rho/dy=\Delta \rho/D$ and $\Delta \rho=-\rho_0 \alpha_c \Delta T$.  The Brunt-Vaisala frequency is related to settling time $\tau$ by $\tau=2\pi/N$.
Assuming a lateral temperature difference of $\Delta T=6000$ K over a width of $D=1000$ km, equation (\ref{N2}) predicts $\tau\approx2$ hours.  Therefore, lateral temperature heterogeneities in the core should homogenize rapidly.  In the solid part of the mantle lateral temperature heterogeneities could last much longer but will not influence the secular cooling rate of the core.


\subsection{Numerical Challenges}
In the current SPH simulations, there are two main challenges concerning the computation of heat. Firstly, SPH can exhibit inaccuracies at the metal-silicate interface, potentially leading to an underestimation of the temperature of core particles at the CMB. Our simulation employed an enhanced density calculation method \cite{reinhardt2017numerical,reinhardt2020bifurcation}. This method  attempts to correct for the errors in the density calculation at the interface, thereby obtaining more accurate values for temperature and pressure, among others.

Secondly, the SPH method does not incorporate heat exchange, meaning metal core particles do not exchange heat with the surrounding mantle. For some scenarios, such as where the core would totally break up first and then disperse into the mantle, the core particles would have extensive contact with the flowing mantle. This may result in an overestimation of the temperature of core particles at the CMB. In the case of direct core-merging collisions, the impactor metal core remains mostly intact and merges with proto-Earth's core during the several hours impact simulation, implying minimal heat exchange with the mantle \cite{zhou2021core}. 

Overall, we expect these numerical issues introduce minimal errors compared to the magnitude of the impact heating and its distribution in the core. Although the above-mentioned issues are worth exploring, they seem unlikely to affect the main conclusions of this study.

\section{Conclusions}\label{conclusions}

In this study we have investigated the core heating of Earth-like planets due to a giant impact during the final stage of accretion. We have considered a range of impact scenarios, varying in impact angle, impact velocity, impactor mass, and initial thermal state. We find that the amount of core heating depends sensitively on the impact conditions.  A grazing impact results in negligible core heating, while an extreme heating of $14,000$ K can occur in a collision with an impact angle of $15^\circ$, impact velocity of $2V_{\text{esc}}$, and an impactor mass of $0.3M_{\oplus}$. The majority of the core heat deposition during the impact is attributed to the merging of the impactor and target cores.

From these simulations we have derived scaling relations for core heating due to a giant impact. The scaling relations predict both the bulk heat deposited into the core and the radial distribution of the impact heating. The heating occurs in two steps, first shock heating followed by core merging. 
 The scaling relation is intended to be general so that it can be applied to a range of impact conditions during accretion. We find the scaling relation predicts core heating most accurately (in a $\chi^2$-sense) for scenarios with substantial core heating, particularly when the average core heating exceeds $\sim$ 2000 K. The radial distribution of impact heat is accurately predicted by the scaling relation for all scenarios with impact angle below $30^\circ$, most impacts at $45^\circ$, and only some with angles greater than $60^\circ$.


Our findings robustly find that giant impacts primarily heat the outer-most core, while the inner-most core experiences a lesser degree of heating. This occurs because much of the impact-related heating comes from the merging of the impactor and target cores.  The impactor core tends to be be less dense and hotter, which prevents mixing of the impactor core deep into the target core because it is thermally stratified.  

Similar giant impact-generated core thermal stratification has been found in previous studies \cite{arkanihamed2010,landeau2016core}.  This impact-generated stratification is expected to impede thermal convection in the core, consequently delaying the generation of a dynamo magnetic field. 
This differs from other studies that conclude that giant impacts can actually help mix the core and instigate dynamo action \cite{reese2010,jacobson2017}.  However, our 3D giant impact models clearly show that the majority of the impact heat is concentrated in the impactor core compared to the proto-Earth core, which will tend to thermally stratify the core after core merging.

Using a parameterized cooling model we estimate the time for the core to cool to an adiabatic state to be $200-700$ Myr, depending on the impact conditions.  For a canonical impact, the estimated cooling time is consistent with ancient paleomagnetic evidence for an active geodynamo $\sim260$ Myr  after the moon-forming giant impact.

\section*{Appendix A}
\setcounter{table}{0} 
\renewcommand{\thetable}{A\arabic{table}} 

In Table \ref{table2}, Cases 1 to 60 include all the simulations performed with the surface temperature set to 2000K. Specifically, Case c represents the canonical impact model.

\begin{longtable}{c l c c | c c c c c}
\caption{Simulation parameters and results summary table}\\
\text{Case} & $\sin(\theta)$ & $M_i/M_{\oplus}$ & $V_i/V_{\text{esc}}$ & $\Delta \bar{T}$ & $\chi^2$  &$M_{\text{planet}}/M_{\oplus}$ & $M_{\text{disc}}/M_{\text{m}}$ & $L_z / L_{\text{EM}}$ \\
\hline
\endfirsthead 

\multicolumn{9}{c}%
{\tablename\ \thetable\ -- \textit{Continued from previous page}} \\
\hline
\text{Case} & $\sin(\theta)$ & $M_i/M_{\oplus}$ & $V_i/V_{\text{esc}}$ & $\Delta \bar{T}$ & $\chi^2$  &$M_{\text{planet}}/M_{\oplus}$ & $M_{\text{disc}}/M_{\text{m}}$ & $L_z / L_{\text{EM}}$ \\
\hline
\endhead 

\hline \multicolumn{9}{r}{\textit{Continued on next page}} \\
\endfoot 

\hline
\multicolumn{9}{l}{\begin{tabular}[t]{@{}p{\textwidth}@{}}$^{a}$Examples marked with an asterisk (*) require two steps to identify their accretion disks. The resolution of Case c is $2\times 10^5$ particles. The vertical line in the middle separates input parameters on the left and output parameters on the right.
\end{tabular}}

\endlastfoot 
1  & 0.2588 & 0.05 & 1 & 3182  & 4.59  & 1.05 & 0.02   & 0.18 \\
2  & 0.2588 & 0.10  & 1 & 6555  & 0.98   & 1.10  & 0.02   & 0.35 \\
3  & 0.2588 & 0.20  & 1 & 9655  & 0.46  & 1.19 & 0.63   & 0.71 \\
4  & 0.2588 & 0.30  & 1 & 11028 & 0.62  & 1.29 & 0.61   & 1.05 \\
5  & 0.2588 & 0.05 & 2 & 6181  & 2.23  & 1.04 & 0.18   & 0.30  \\
6  & 0.2588 & 0.10  & 2 & 8280  & 1.01  & 1.07 & 0.64   & 0.52 \\
7  & 0.2588 & 0.20  & 2 & 11748 & 0.61  & 1.10  & 1.72   & 0.78 \\
8  & 0.2588 & 0.30  & 2 & 14068 & 0.17  & 1.13 & 8.53   & 1.04 \\
9  & 0.2588 & 0.05 & 3 & 7835  & 0.52  & 1.01 & 0.21   & 0.32 \\
10 & 0.2588 & 0.10  & 3 & 9896  & 0.17  & 0.95 & 2.79   & 0.33 \\
11 & 0.2588 & 0.20  & 3 & 10884 & 0.06  & 0.84 & 4.45   & 0.26 \\
*12 & 0.2588 & 0.30  & 3 & 10814 & 1.89  & 0.73 & 0.00   & 0.25  \\
13 & 0.5000 & 0.05 & 1 & 2451  & 6.30 & 1.05 & 0.08   & 0.32 \\
14 & 0.5000 & 0.10  & 1 & 4208  & 4.06  & 1.09 & 1.40  & 0.63 \\
15 & 0.5000 & 0.20  & 1 & 6383  & 2.76 & 1.18 & 5.53   & 1.29 \\
16 & 0.5000 & 0.30 & 1 & 8553  & 1.37  & 1.29 & 17.74  & 2.05 \\
17 & 0.5000 & 0.05 & 2 & 3146  & 2.73  & 1.02 & 1.65   & 0.40  \\
18 & 0.5000 & 0.10  & 2 & 3982  & 2.08  & 1.03 & 7.00   & 0.55 \\
19 & 0.5000 & 0.20  & 2 & 4735  & 2.82  & 1.00 & 1.54   & 0.45 \\
20 & 0.5000 & 0.30  & 2 & 5497  & 2.59  & 0.99 & 2.28   & 0.43 \\
21 & 0.5000 & 0.05 & 3 & 4223  & 1.24  & 0.98 & 1.40 & 0.29 \\
22 & 0.5000 & 0.10  & 3 & 5627  & 0.79  & 0.96 & 1.24   & 0.31 \\
23 & 0.5000 & 0.20  & 3 & 7174  & 0.64  & 0.93 & 1.68   & 0.31 \\
24 & 0.5000 & 0.30  & 3 & 8111  & 0.56  & 0.91 & 2.47   & 0.31 \\
25 & 0.7071 & 0.05 & 1 & 1537  & 4.7  & 1.04 & 0.23   & 0.40  \\
26 & 0.7071 & 0.10  & 1 & 2498  & 3.44  & 1.08 & 5.98   & 0.79 \\
27 & 0.7071 & 0.20  & 1 & 4719  & 4.69  & 1.16 & 69.64  & 1.60  \\
28 & 0.7071 & 0.30  & 1 & 6242  & 2.53  & 1.25 & 211.55 & 2.90  \\
29 & 0.7071 & 0.05 & 2 & 1136  & 2.85  & 1.01 & 0.16   & 0.20  \\
30 & 0.7071 & 0.10  & 2 & 1617  & 2.78  & 1.01 & 0.73   & 0.25 \\
31 & 0.7071 & 0.20  & 2 & 2276  & 2.92  & 1.00  & 0.01   & 0.32 \\
*32 & 0.7071 & 0.30  & 2 & 2717  & 2.67  & 1.00 & 0.00  & 0.37 \\
33 & 0.7071 & 0.05 & 3 & 1821  & 2.41  & 1.00  & 0.02   & 0.16 \\
34 & 0.7071 & 0.10  & 3 & 2460  & 2.62  & 0.99 & 0.18   & 0.21 \\
*35 & 0.7071 & 0.20  & 3 & 3203  & 1.40  & 0.99 &  0.00  & 0.27  \\
*36 & 0.7071 & 0.30  & 3 & 3757  & 1.94  & 0.98 &  0.00  & 0.30 \\
37 & 0.8660 & 0.05 & 1 & 709   & 5.41  & 1.03 & 0.19   & 0.30 \\
38 & 0.8660 & 0.10  & 1 & 1007  & 4.37  & 1.04 & 2.45   & 0.37 \\
39 & 0.8660 & 0.20  & 1 & 1256  & 4.99  & 1.04 & 0.49   & 0.40  \\
40 & 0.8660 & 0.30  & 1 & 1437  & 4.29  & 1.23 & 362.13 & 4.42 \\
41 & 0.8660 & 0.05 & 2 & 247   & 4.30  & 1.00  & 1.05   & 0.08 \\
42 & 0.8660 & 0.10  & 2 & 358   & 3.24  & 1.00  & 0.18   & 0.10  \\
*43 & 0.8660 & 0.20  & 2 & 519   & 2.08  & 1.00  & 0.00   & 0.14 \\
*44 & 0.8660 & 0.30  & 2 & 651   & 2.59  & 1.00  &  0.00  &  0.17 \\
45 & 0.8660 & 0.05 & 3 & 337   & 4.33  & 1.00  & 0.04   & 0.06 \\
*46 & 0.8660 & 0.10  & 3 & 476   & 2.77  & 1.00 & 0.00  &  0.08 \\
*47 & 0.8660 & 0.20  & 3 & 471   & 1.56  & 1.00 &  0.00  &  0.10 \\
*48 & 0.8660 & 0.30  & 3 & 815   & 1.38  & 1.00 &  0.00  &  0.12 \\
49 & 0.9659 & 0.05 & 1 & 220   & 8.18  & 1.01 & 0.05   & 1.15 \\
50 & 0.9659 & 0.10  & 1 & 327   & 3.50  & 1.02 & 0.13   & 1.20  \\
51 & 0.9659 & 0.20  & 1 & 564   & 3.57  & 1.02 & 0.13   & 0.26 \\
52 & 0.9659 & 0.30  & 1 & 814   & 4.59  & 1.02 & 0.13   & 0.31 \\
53 & 0.9659 & 0.05 & 2 & 40    & 4.03 & 1.00    & 0.22   & 0.02 \\
54 & 0.9659 & 0.10  & 2 & 71    & 5.4  & 1.00   & 0.16   & 0.03 \\
*55 & 0.9659 & 0.20  & 2 & 132   & 3.8  & 1.00 & 0.00   & 0.05 \\
*56 & 0.9659 & 0.30  & 2 & 197   & 3.76  & 1.00  & 0.00  & 0.06  \\
57 & 0.9659 & 0.05 & 3 & 32    & 30.65  & 1.00 & 0.00    & 0.01 \\
*58 & 0.9659 & 0.10  & 3 & 50    & 7.33  & 1.00  & 0.00  & 0.02  \\
*59 & 0.9659 & 0.20  & 3 & 84    & 4.02  &  1.00  & 0.00   & 0.02 \\
*60 & 0.9659 & 0.30  & 3 & 120   & 5.24  &   1.00 & 0.00  & 0.03 \\
c  & 0.7300 & 0.15  & 1 & 3099   & 3.51  &   0.99 & 1.65  & 1.06 
\label{table2}
\end{longtable}

In \textit{Table \ref{table2}}, $ M_{\text{planet}} $ denotes the mass of the proto-Earth post-collision, $ M_{\text{disc}} $ signifies the mass of the accretion disc formed following the collision, and $ M_m $ corresponds to the present-day mass of the Moon. Additionally, $ L_z $ represents the combined angular momentum of the accretion disc and the proto-Earth after the collision, while $ L_{\text{EM}} $ encapsulates the total angular momentum of the current Earth-Moon system. Other parameter symbols have been explained in the previous sections.

Examples marked with an asterisk (*) cannot have their accretion disks directly identified using the method described by \citeA{canup2001scaling}. We first use the clump finder method which employs a friends-of-friends algorithm to determine whether a given SPH particle belongs to a gravity-bound clump \cite{stadel2001cosmological,reinhardt2022forming}.

\acknowledgments
 
We thank the anonymous reviewer for their constructive comments, which have improved the quality of this paper. We thank Prof. ZhiXue Du for the discussion that inspired us. Thanks to Dr. Zhongtian Zhang for the valuable suggestions on data fitting. Special thanks to Prof. Yun Liu for supporting You Zhou's participation in this project. 

\section*{Open Research}

The numerical model is composed of three primary components: the Gasoline source code, EOS interfaces, and the Equations of State source code. All of these are included in the OSF data repository for this paper in \cite{Zhou2024}.
The results and plotting scripts are also available in \cite{Zhou2024}. 
\section*{Author Contribution}
Conceptualization: You Zhou, Peter E. Driscoll
Data curation: You Zhou, Peter E. Driscoll
Formal analysis: You Zhou, Peter E. Driscoll, Mingming Zhang, Christian Reinhardt
Methodology: You Zhou, Peter E. Driscoll
Resources: You Zhou, Peter E. Driscoll
Software: You Zhou, Christian Reinhardt, Thomas Meier,  Peter E. Driscoll
Validation: You Zhou, Peter E. Driscoll
Visualization: You Zhou, Peter E. Driscoll
Writing – original draft: You Zhou, Peter E. Driscoll
Writing – review \& editing: You Zhou, Peter E. Driscoll, Christian Reinhardt


%
%


%
%
%
%
%

\bibliography{reference}      

\end{document}


%
%


\title{Supporting Information for "A scaling relation for core heating by giant impacts and implications for dynamo onset"}
%
%

%
%



\authors{You Zhou\affil{1,2}, Peter E. Driscoll\affil{2}, Mingming Zhang\affil{2}, Christian Reinhardt\affil{3,4},Thomas Meier\affil{3}}

\affiliation{1}{Planetary Science Research Center, College of Earth Sciences, Chengdu University of Technology, Chengdu, SiChuan, China}
\affiliation{2}{Earth and Planets Laboratory, Carnegie Institution for Science, Washington, DC, USA}
\affiliation{3}{Department of Astrophysics, University of Zurich, Winterthurerstrasse 190, CH-8057 Zurich, Switzerland}
\affiliation{4}{Physics Institute, Space Research and Planetary Sciences, University of Bern, Sidlerstrasse 5, CH-3012 Bern, Switzerland}

%
%

%

\begin{article}

%
%

\noindent\textbf{Contents of this file}
\begin{enumerate}
\item Figures S1 to S10
\end{enumerate}

\end{article}
\clearpage

\begin{figure}
\noindent\includegraphics[width=\textwidth]{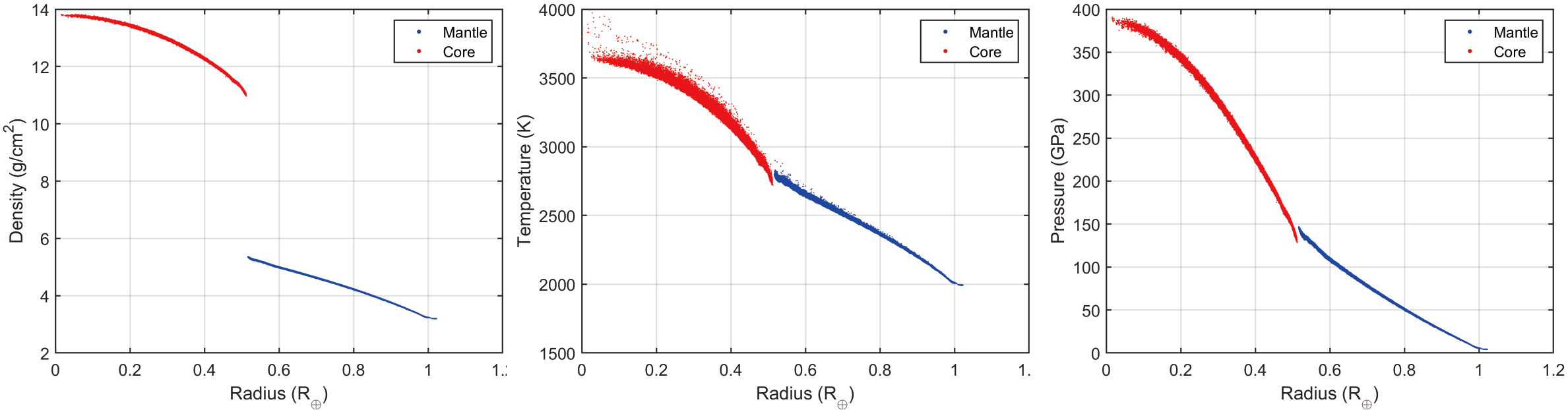}
\caption{Simulation using low-noise initial conditions, illustrating the primitive Earth's initial structure in density, temperature, and pressure, with a surface temperature set at 2000K. The left panel depicts the density variation with radius, the middle panel represents the temperature variation with radius, and the right panel showcases the pressure variation with radius.}
\label{figuresample10}
\end{figure}

\begin{figure}
\noindent\includegraphics[width=\textwidth]{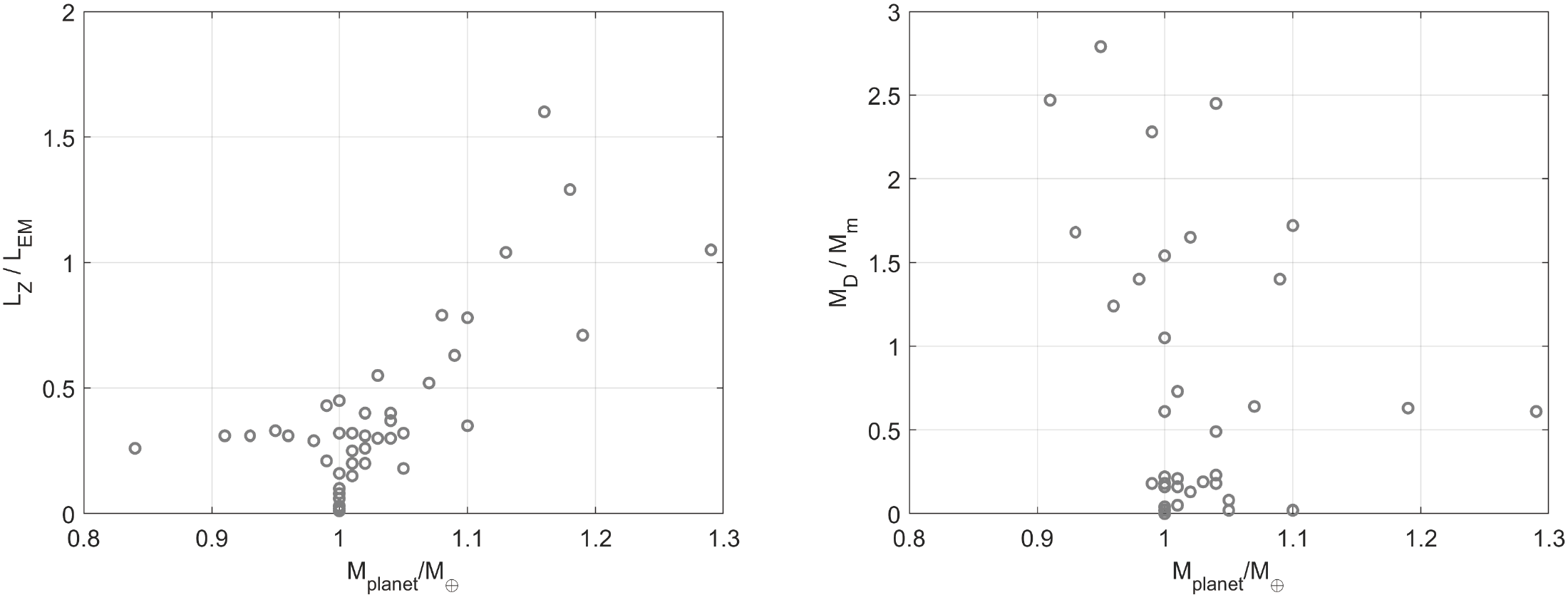}
\caption{The left panel illustrates the relationship between the planet's mass after collision and the angular momentum of the planet and the accretion disk, while the right panel depicts the relationship between the planet's mass after collision and the mass of the accretion disk.}
\label{figuresample8}
\end{figure}

\begin{figure}
\noindent\includegraphics[width=\textwidth]{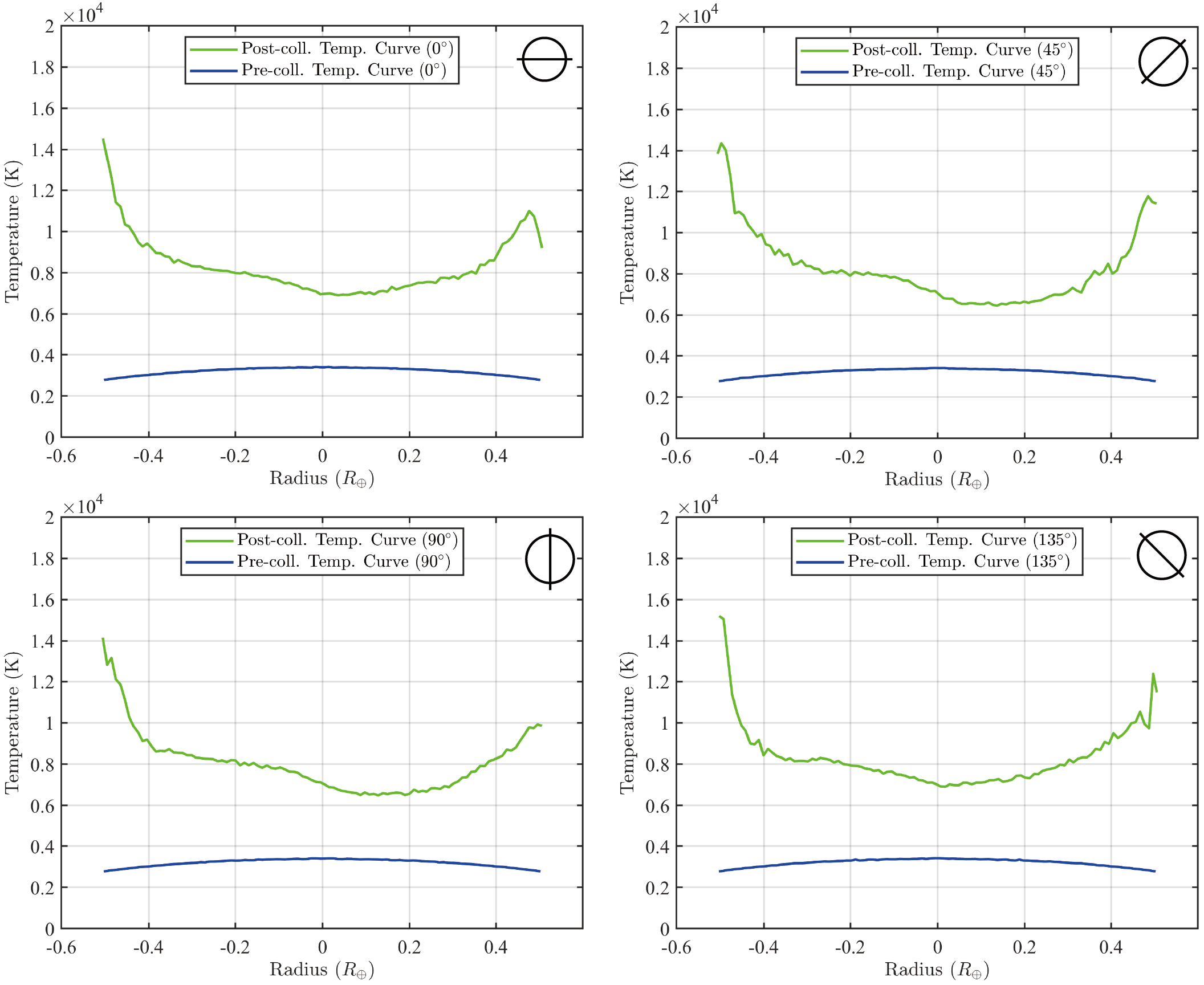}
\caption{The post-impact temperature profiles of the core in four directions are shown, corresponding to angles of 0\textdegree, 45\textdegree, 90\textdegree, and 135\textdegree\ relative to the X-axis. In this instance, the impactor's mass is 0.1M$_{\oplus}$, the impact velocity is 1V$_{\text{esc}}$, and the impact angle is 15\textdegree. Blue lines indicate the temperature profiles along these directions before the impact, while green lines represent these profiles after the impact. The values of the temperature profiles are the average temperatures of particles within a 0.1R$_{\oplus}$ radius range around each profile line. The legends in the top right corners of each panel denote the direction of each temperature profile. The procedure for identifying the core-mantle boundary is detailed in the Methods section (section 2).}
\label{15du}
\end{figure}

\begin{figure}
\noindent\includegraphics[width=\textwidth]{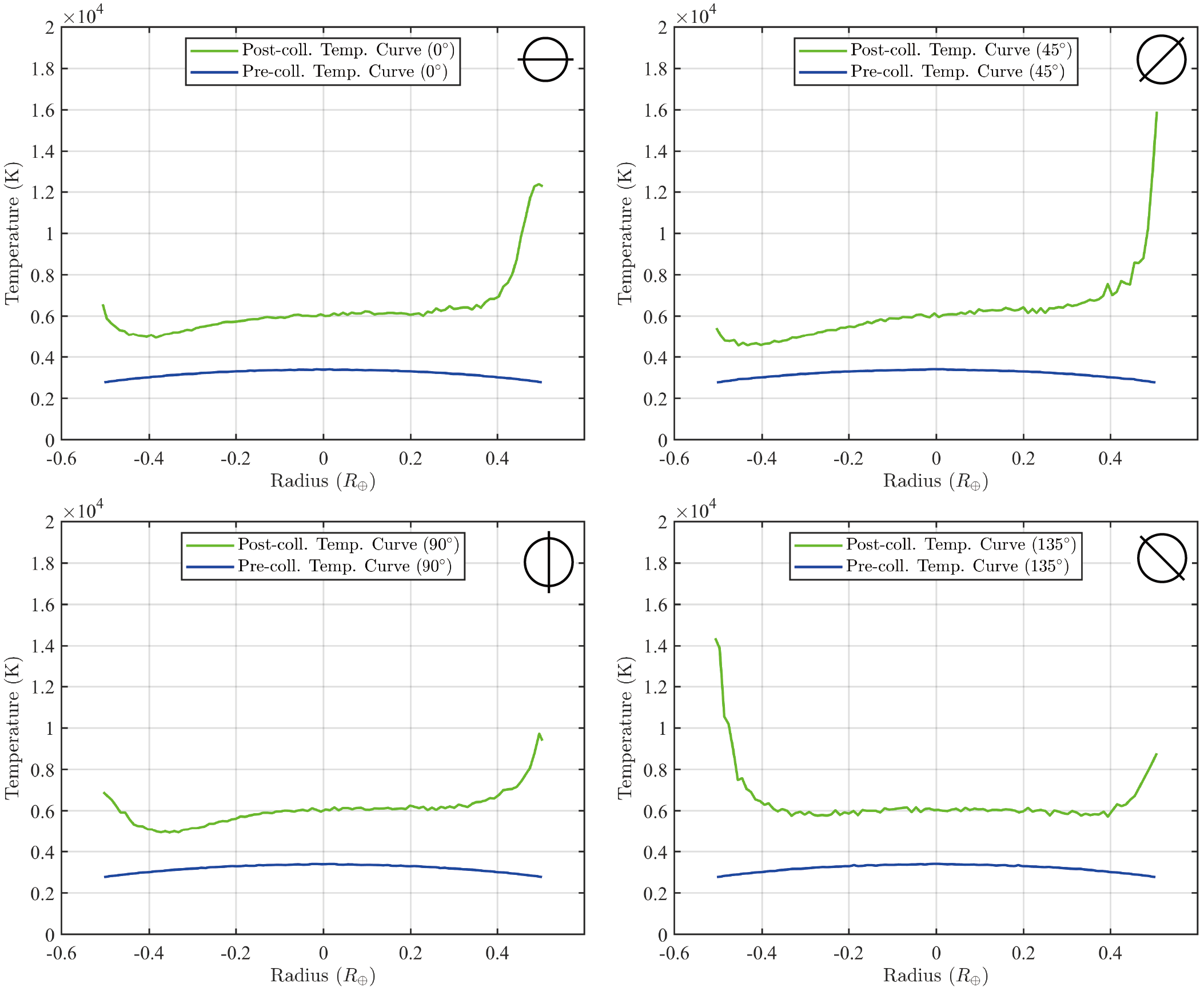}
\caption{The post-impact temperature profiles of the core in four directions are shown, corresponding to angles of 0\textdegree, 45\textdegree, 90\textdegree, and 135\textdegree\ relative to the X-axis. In this instance, the impactor's mass is 0.1M$_{\oplus}$, the impact velocity is 1V$_{\text{esc}}$, and the impact angle is 30\textdegree. Blue lines indicate the temperature profiles along these directions before the impact, while green lines represent these profiles after the impact. The values of the temperature profiles are the average temperatures of particles within a 0.1R$_{\oplus}$ radius range around each profile line. The legends in the top right corners of each panel denote the direction of each temperature profile. The procedure for identifying the core-mantle boundary is detailed in the Methods section (section 2).}
\label{30du}
\end{figure}

\begin{figure}
\noindent\includegraphics[width=\textwidth]{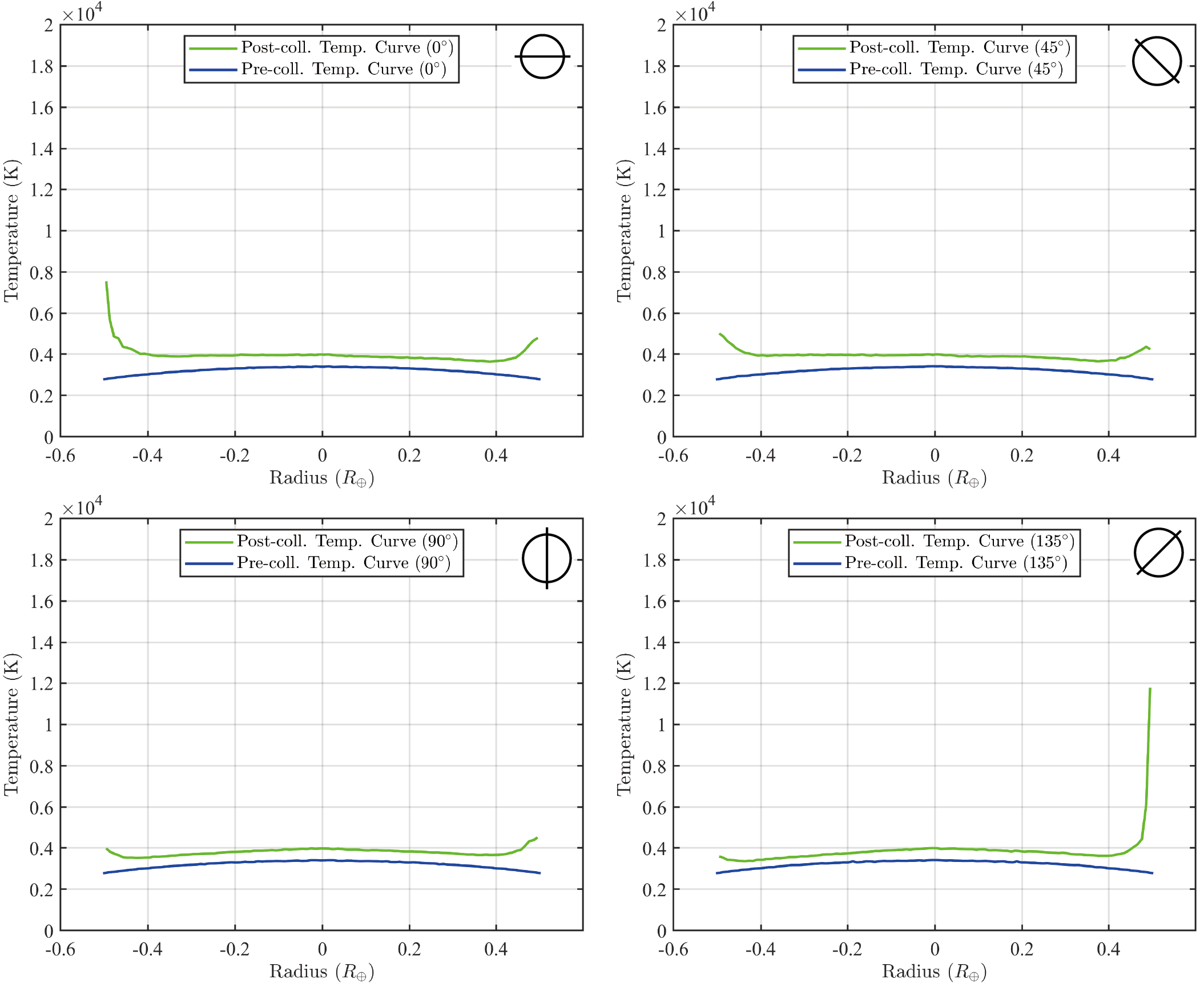}
\caption{The post-impact temperature profiles of the core in four directions are shown, corresponding to angles of 0\textdegree, 45\textdegree, 90\textdegree, and 135\textdegree\ relative to the X-axis. In this instance, the impactor's mass is 0.1M$_{\oplus}$, the impact velocity is 1V$_{\text{esc}}$, and the impact angle is 60\textdegree. Blue lines indicate the temperature profiles along these directions before the impact, while green lines represent these profiles after the impact. The values of the temperature profiles are the average temperatures of particles within a 0.1R$_{\oplus}$ radius range around each profile line. The legends in the top right corners of each panel denote the direction of each temperature profile. The procedure for identifying the core-mantle boundary is detailed in the Methods section (section 2).}
\label{60du}
\end{figure}

\begin{figure}
\noindent\includegraphics[width=\textwidth]{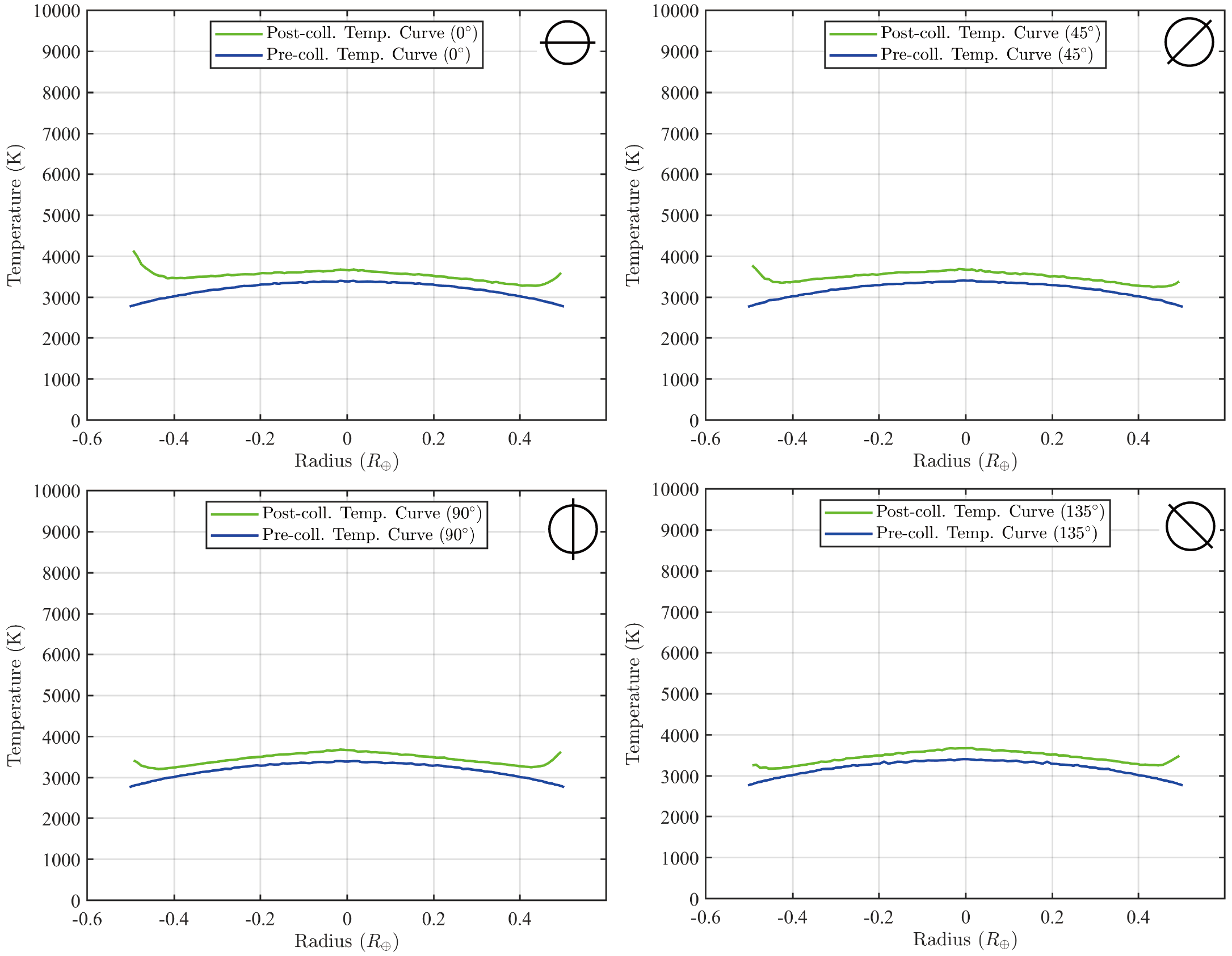}
\caption{The post-impact temperature profiles of the core in four directions are shown, corresponding to angles of 0\textdegree, 45\textdegree, 90\textdegree, and 135\textdegree\ relative to the X-axis. In this instance, the impactor's mass is 0.1M$_{\oplus}$, the impact velocity is 1V$_{\text{esc}}$, and the impact angle is 75\textdegree. Blue lines indicate the temperature profiles along these directions before the impact, while green lines represent these profiles after the impact. The values of the temperature profiles are the average temperatures of particles within a 0.1R$_{\oplus}$ radius range around each profile line. The legends in the top right corners of each panel denote the direction of each temperature profile. The procedure for identifying the core-mantle boundary is detailed in the Methods section (section 2).}
\label{75du}
\end{figure}

\begin{figure}
\noindent\includegraphics[width=\textwidth]{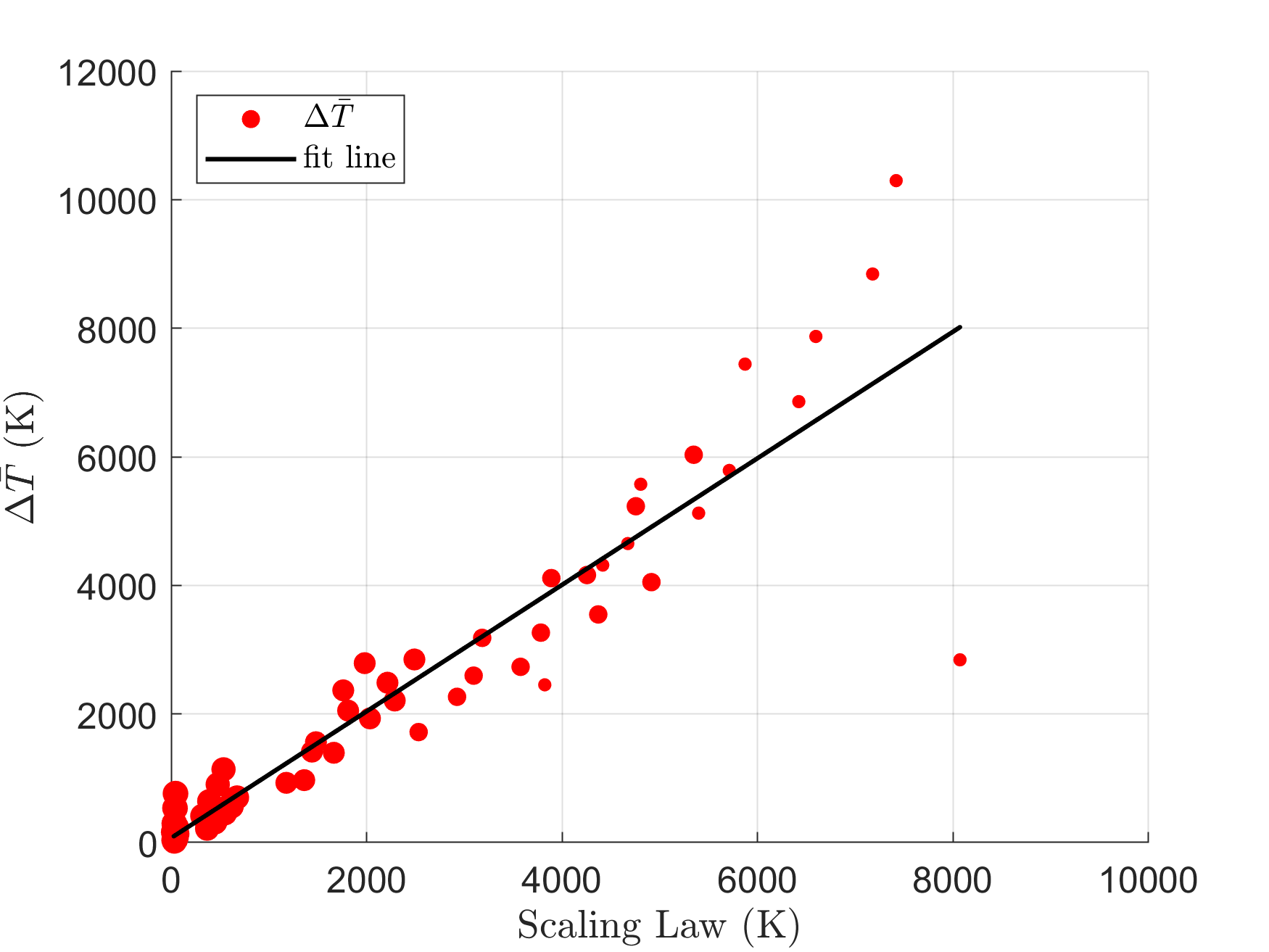}
\caption{The comparison of the fitted values of \(\Delta \bar{T}\) with the actual values of \(\Delta \bar{T}\), using the scaling relation from Equation~(25). The Root Mean Square Error (RMSE) of this fit is 939.1619. Each dot's size is multiplied by the sine of its impact angle, indicating that smaller dots correspond to lower impact angles (more vertical).
}
\label{Tbar}
\end{figure}

\begin{figure}
\noindent\includegraphics[width=\textwidth]{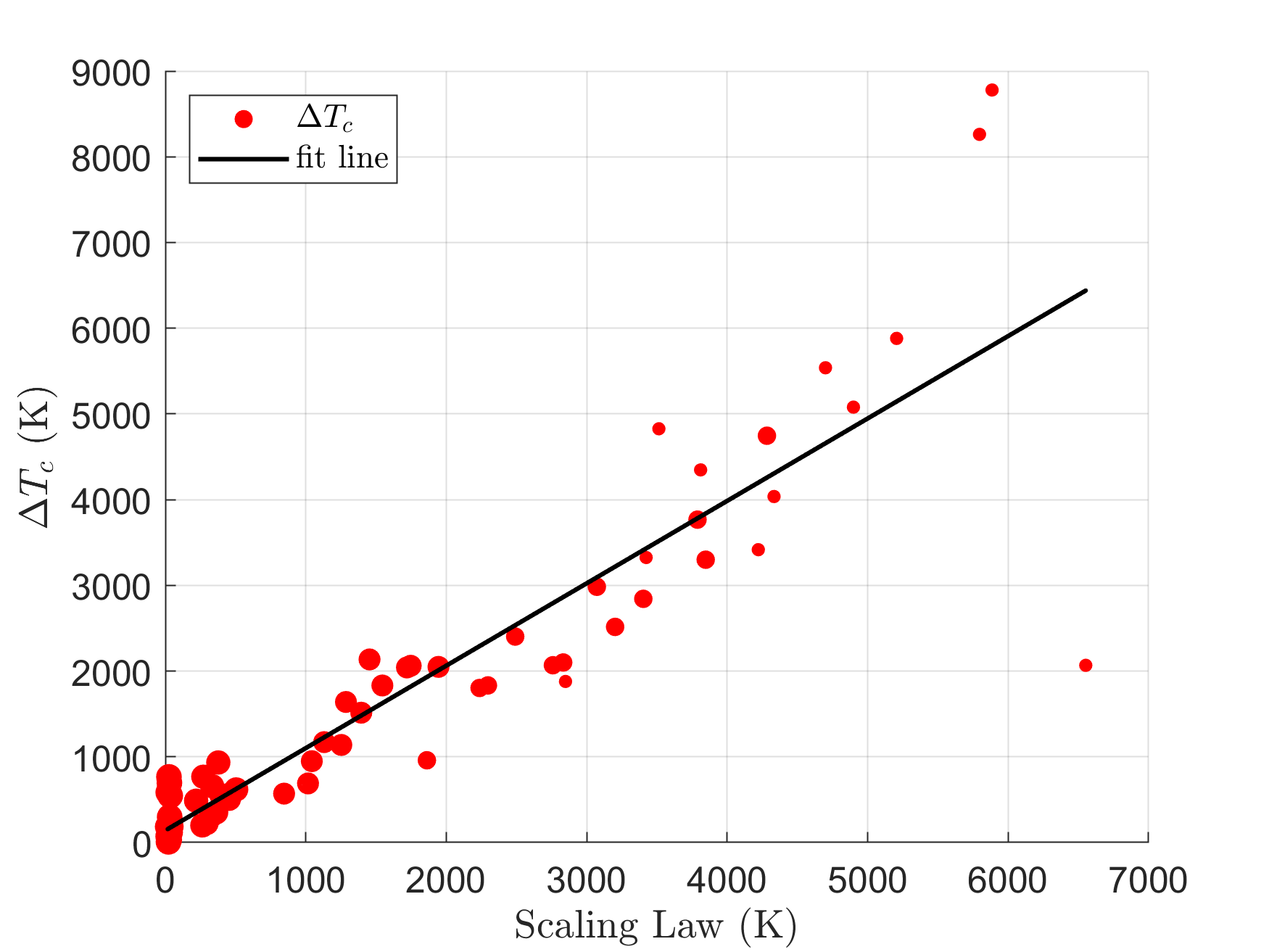}
\caption{The comparison of the fitted values of $\Delta T_c$ with the actual values of $\Delta T_c$, using the scaling relation from Equation~(25). The Root Mean Square Error (RMSE) of this fit is 881.3975. Each dot's size is multiplied by the sine of its impact angle, indicating that smaller dots correspond to lower impact angles (more vertical).}
\label{Tc}
\end{figure}

\begin{figure}
\noindent\includegraphics[width=\textwidth]{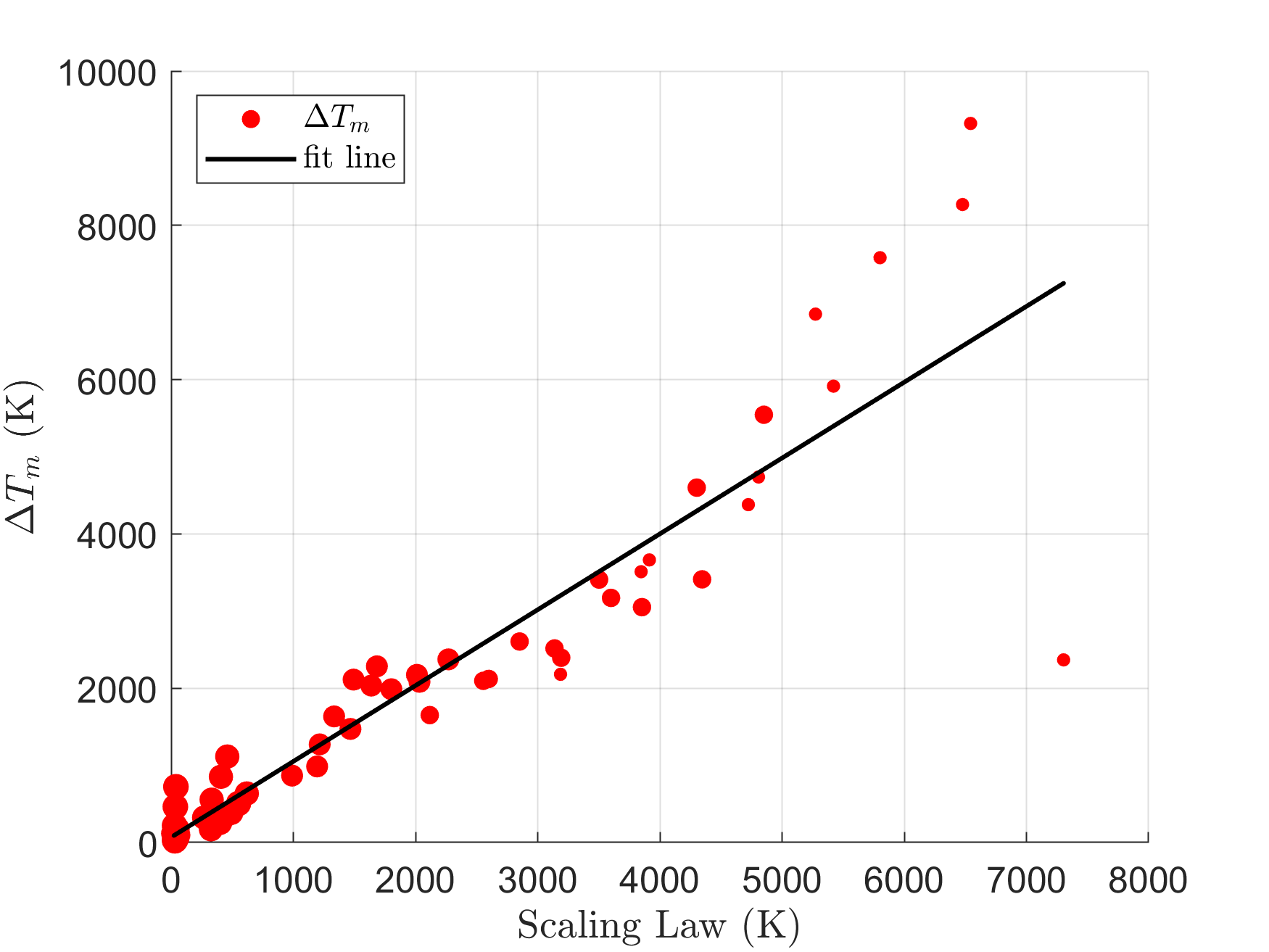}
\caption{The comparison of the fitted values of $\Delta T_m$ with the actual values of $\Delta T_m$, using the scaling relation from Equation~(25). The Root Mean Square Error (RMSE) of this fit is 905.9080. Each dot's size is multiplied by the sine of its impact angle, indicating that smaller dots correspond to lower impact angles (more vertical).}
\label{Th}
\end{figure}

\begin{figure}
\noindent\includegraphics[width=\textwidth]{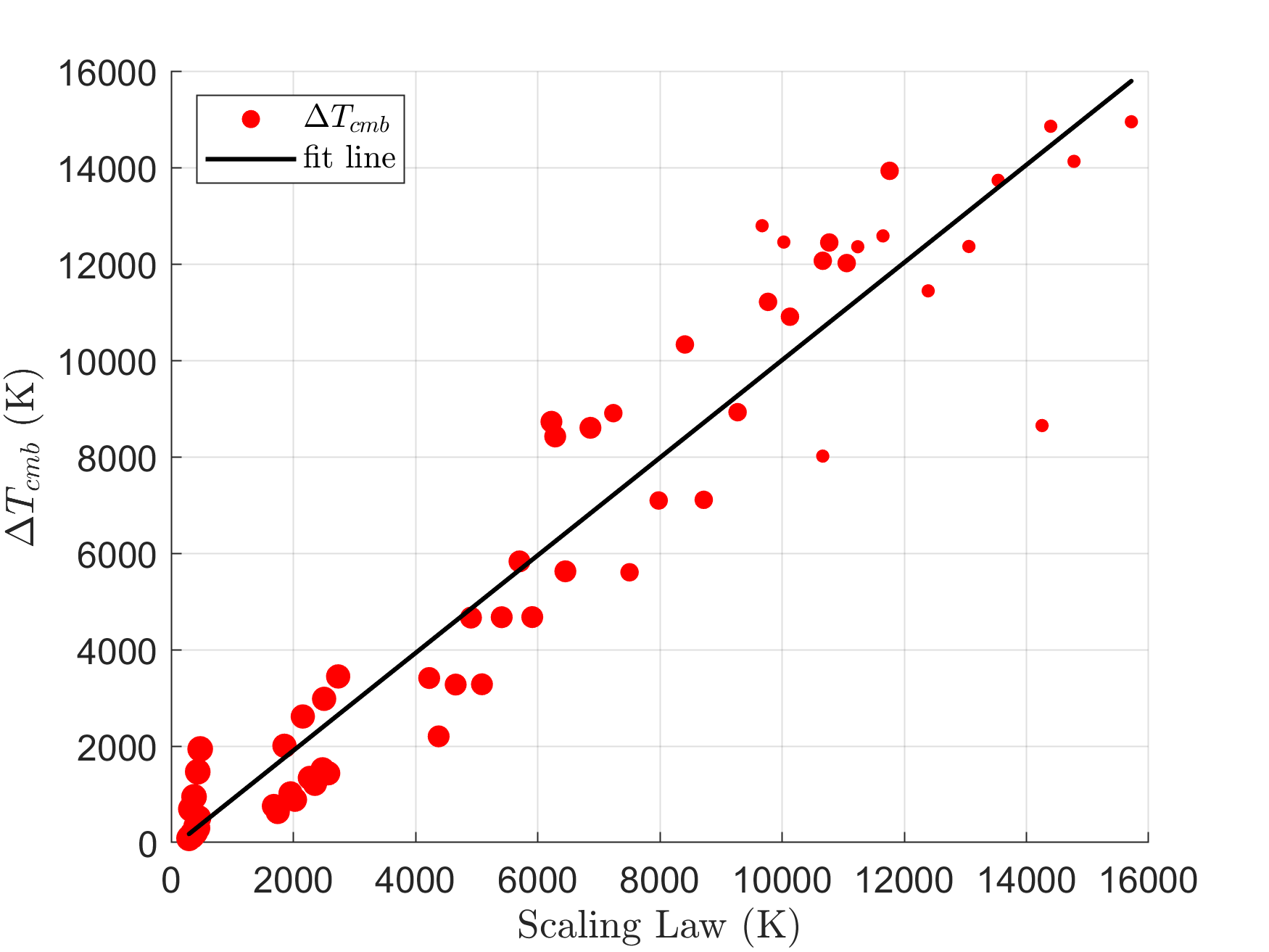}
\caption{The comparison of the fitted values of $\Delta T_{cmb}$ with the actual values of $\Delta T_{cmb}$, using the scaling relation from Equation~(25). The Root Mean Square Error (RMSE) of this fit is 1450.4. Each dot's size is multiplied by the sine of its impact angle, indicating that smaller dots correspond to lower impact angles (more vertical).}
\label{Thr}
\end{figure}